\documentclass[useAMS,usenatbib]{mn2e}

\usepackage{epsfig}
\usepackage{amsfonts}
\usepackage[usenames]{color}
\usepackage[usenames,dvipsnames]{xcolor}
\usepackage{url}
\usepackage{subfigure}
\usepackage{times,graphicx,amsmath,amsfonts,amssymb,aas_macros,epstopdf,hyperref}
\usepackage[normalem]{ulem}
\usepackage[T1]{fontenc}
\usepackage{multirow}
\usepackage{aecompl}

\usepackage{hyperref}
\hypersetup{
    colorlinks=true,
    linkcolor=blue,
    citecolor=blue,
}

\def\etal{{\frenchspacing\it et al.}}
\def\ie{{\frenchspacing\it i.e.}}
\def\eg{{\frenchspacing\it e.g.}}

\def\be{\begin{equation}}
\def\ee{\end{equation}}
\def\ba{\begin{eqnarray}}
\def\ea{\end{eqnarray}}

\newcommand{\hompc}{\,h\,{\rm Mpc}^{-1}}
\newcommand{\mpcoh}{\,h^{-1}\,{\rm Mpc}}

\def\k{{\bf k}}

\def\LaTeX{L\kern-.36em\raise.3ex\hbox{a}\kern-.15em
    T\kern-.1667em\lower.7ex\hbox{E}\kern-.125emX}

\begin{document}

\voffset-1.25cm
\title[Tomographic BAO analysis of BOSS DR12 sample]{The clustering of galaxies in the completed SDSS-III Baryon Oscillation Spectroscopic Survey: tomographic BAO analysis of DR12 combined sample in configuration space}
\author[Wang \etal]{
\parbox{\textwidth}{
Yuting Wang$^{1,2}$\thanks{Email: ytwang@nao.cas.cn}, Gong-Bo Zhao$^{1,2}$\thanks{Email: gbzhao@nao.cas.cn}, Chia-Hsun Chuang$^{3,4}$, Ashley J. Ross$^{5}$, Will J. Percival$^{2}$, H\'ector Gil-Mar\'{\i}n$^{2}$, Antonio J. Cuesta$^{6}$, Francisco-Shu Kitaura$^{4}$, Sergio Rodriguez-Torres$^{3,7,8}$, Joel R. Brownstein$^{9}$, Daniel J. Eisenstein$^{10}$, Shirley Ho$^{11,12,13}$, Jean-Paul Kneib$^{14}$, Matthew D. Olmstead$^{15}$, Francisco Prada$^{7}$, Graziano Rossi$^{16}$, Ariel G. S\'anchez$^{17}$, Salvador Salazar-Albornoz$^{18,17}$, Daniel Thomas$^{2}$, Jeremy Tinker$^{19}$, Rita Tojeiro$^{20}$, Mariana Vargas-Maga\~{n}a$^{21}$, Fangzhou Zhu$^{22}$}
\vspace*{15pt} \\
$^{1}$ National Astronomy Observatories, Chinese Academy of Science, Beijing, 100012, P. R. China\\
$^{2}$ Institute of Cosmology \& Gravitation, University of Portsmouth, Dennis Sciama Building, Portsmouth, PO1 3FX, UK\\
$^{3}$ Instituto de F\'{\i}sica Te\'orica, (UAM/CSIC), Universidad Aut\'onoma de Madrid,  Cantoblanco, E-28049 Madrid, Spain \\
$^{4}$ Leibniz-Institut f\"ur Astrophysik Potsdam (AIP), An der Sternwarte 16, 14482 Potsdam, Germany \\
$^{5}$ Center for Cosmology and AstroParticle Physics, The Ohio State University, Columbus, OH 43210, USA \\
$^{6}$ Institut de Ci\`encies del Cosmos (ICCUB), Universitat de Barcelona (IEEC- UB), Mart\'{\i} i Franqu\`es 1, E-08028 Barcelona, Spain \\
$^{7}$ Campus of International Excellence UAM+CSIC, Cantoblanco, E-28049 Madrid, Spain \\
$^{8}$ Departamento de F\'{\i}sica Te\'orica, Universidad Aut\'onoma de Madrid, Cantoblanco, E-28049, Madrid, Spain\\
$^{9}$ Department of Physics and Astronomy, University of Utah, 115 S 1400 E, Salt Lake City, UT 84112, USA \\
$^{10}$ Harvard-Smithsonian Center for Astrophysics, 60 Garden St., Cambridge, MA 02138, USA \\
$^{11}$ McWilliams Center for Cosmology, Department of Physics, Carnegie Mellon University, 5000 Forbes Ave., Pittsburgh, PA 15213\\
$^{12}$ Lawrence Berkeley National Laboratory, 1 Cyclotron Rd, Berkeley, CA 94720 \\
$^{13}$ Department of Physics, University of California, Berkeley, CA 94720 \\
$^{14}$ Laboratoire d'Astrophysique, Ecole Polytechnique F\'ed\'erale de Lausanne (EPFL), Observatoire de Sauverny, CH-1290 Versoix, Switzerland \\
$^{15}$ Department of Chemistry and Physics, King's College, 133 North River St, Wilkes Barre, PA 18711, USA \\
$^{16}$ Department of Astronomy and Space Science, Sejong University, Seoul 143-747, Korea \\
$^{17}$ Max-Planck-Institut f\"ur extraterrestrische Physik, Postfach 1312, Giessenbachstr., 85741 Garching, Germany \\
$^{18}$ Universit\"at Sternwarte Mu\"anchen, Ludwig Maximilian Universit\"at, Munich, Germany \\
$^{19}$ Center for Cosmology and Particle Physics, Department of Physics, New York University, 4 Washington Place, New York, NY 10003, USA \\
$^{20}$ School of Physics and Astronomy, University of St Andrews, North Haugh, St Andrews KY16 9SS, UK \\
$^{21}$ Instituto de Fisica, Universidad Nacional Autonoma de Mexico, Apdo. Postal 20-364, Mexico \\
$^{22}$ Department of Physics, Yale University, New Haven, CT 06511, USA 
}
\date{\today} 
\pagerange{\pageref{firstpage}--\pageref{lastpage}}

\label{firstpage}

\maketitle

\clearpage

\begin{abstract} 

We perform a tomographic baryon acoustic oscillations analysis using the two-point galaxy correlation function measured from the combined sample of BOSS DR12, which covers the redshift range of $0.2<z<0.75$. Splitting the sample into multiple overlapping redshift slices to extract the redshift information of galaxy clustering, we obtain a measurement of $D_A(z)/r_d$ and $H(z)r_d$ at nine effective redshifts with the full covariance matrix calibrated using MultiDark-Patchy mock catalogues. Using the reconstructed galaxy catalogues, we obtain the precision of $1.3\%-2.2\%$ for $D_A(z)/r_d$ and $2.1\%-6.0\%$ for $H(z)r_d$. To quantify the gain from the tomographic information, we compare the constraints on the cosmological parameters using our 9-bin BAO measurements, the consensus 3-bin BAO and RSD measurements at three effective redshifts in \citet{Alam2016}, and the non-tomographic (1-bin) BAO measurement at a single effective redshift. Comparing the 9-bin with 1-bin constraint result, it can improve the dark energy Figure of Merit by a factor of 1.24 for the Chevallier-Polarski-Linder parametrisation for equation of state parameter $w_{\rm DE}$. The errors of $w_0$ and $w_a$ from 9-bin constraints are slightly improved when compared to the 3-bin constraint result. 

\end{abstract}


\section{Introduction}
\label{sec:intro}
The accelerating expansion of the Universe was discovered by the observation of type Ia supernovae \citep{Riess,Perlmutter}. Understanding the physics of the cosmic acceleration is one of the major challenges in cosmology. In the framework of general relatively (GR), a new energy component with a negative pressure, dubbed dark energy (DE), can be the source driving the cosmic acceleration. Observations reveal that the DE component dominates the current Universe \citep{DEreview}. However, the nature of DE remains unknown. Large cosmological surveys, especially for galaxy redshift surveys, can provide key observational support for the study of DE. 

Galaxy redshift surveys are used to map the large scale structure of the Universe, and extract the signal of baryon acoustic oscillation (BAO). The BAO, produced by the competition between gravity and radiation due to the couplings between baryons and photons before the cosmic recombination, leave an imprint on the distribution of galaxies at late times. After the photons decouple, the acoustic oscillations are frozen and correspond to a characteristic scale, determined by the comoving sound horizon at the drag epoch, $r_d\sim150\,\rm Mpc$. This feature corresponds to an excess on the 2-point correlation function, or a series of wiggles on the power spectrum. The acoustic scale is regarded as a standard ruler to measure the cosmic expansion history, and to constrain cosmological parameters \citep{Eisenstein2005}. If assuming an isotropic galaxies clustering, the combined volume distance, $D_V(z) \equiv \left[cz (1+z)^2 D_A(z)^2 H^{-1}(z) \right]^{1/3}$, where $H(z)$ is the Hubble parameter and $D_A(z)$ is the angular diameter distance, can be measured using the angle-averaged 2-point correlation function, $\xi_0(s)$ \citep{Eisenstein2005,Kazin2010,Beutler2011,Blake2011} or power spectrum $P_0(k)$ \citep{Tegmark2006,Percival2007,Reid2010}. However, in principle the clustering of galaxies is anisotropic, the BAO scale can be measured in the radial and transverse directions to provide the Hubble parameter, $H(z)$, and angular diameter distance, $D_A(z)$, respectively. As proposed by \citealt{Padmanabhan2008}, the ``multipole'' projection of the full 2D measurement of power spectrum, $P_{\ell}(k)$, were used to break the degeneracy of $H(z)$ and $D_A(z)$.  This multipole method was applied into the correlation function \citep{Chuang2012, Chuang2013, Xu2013}. Alternative ``wedge'' projection of correlation function, $\xi_{\Delta \mu}(s)$, was used to constrain parameters, $H(z)$ and $D_A(z)$ \citep{Kazin2012, Kazin2013}. In \citet{Anderson2014}, the anisotropic BAO analysis was performed using these two projections of correlation function from SDSS-III Baryon Oscillation Spectroscopic Survey (BOSS) DR10 and DR11 samples.

The BOSS \citep{Dawson2013}, which is part of SDSS-III \citep{Eisenstein2011}, has provided the Data Release 12 \citep{Alam}. With a redshift cut, the whole samples are split into the `low-redshift' samples (LOWZ) in the redshift range $0.15<z<0.43$ and `constant stellar mass' samples (CMASS) in the redshift range $0.43<z<0.7$. Using these catalogues, the BAO peak position was measured at two effective redshifts, $z_{\rm eff}=0.32$ and $z_{\rm eff}=0.57$, in the multipoles of correlation function \citep{Cuesta} or power spectrum \citep{Gil-Mar}. \citealt{CF_4bins} proposed to divide each sample of LOWZ and CMASS into two independent redshift bins, thus to test the extraction of redshift information from galaxy clustering. They performed the measurements on BAO and growth rate at four effective redshifts, $z_{\rm eff}=0.24,\,0.37,\,0.49$ and $0.64$ \citep{CF_4bins}.

The completed data release of BOSS will provide a combined sample, covering the redshift range from $0.2$ to $0.75$.  The sample is divided into three redshift bins, \ie, two independent redshift bins, $0.2<z<0.5$ and $0.5<z<0.75$, and an overlapping redshift bin, $0.4<z<0.6$. The BAO signal is measured at the three effective redshifts, $z_{\rm eff}=0.38, \, 0.51$ and $0.61$ using the configuration-space correlation function \citep{CF-sysweight, Mariana2016} or Fourier-space power spectrum \citep{pk-BAO}. 

As the tomographic information of galaxy clustering is important to constrain the property of DE \citep{Albornoz2014,DE-recon}, we will extract the information of redshift evolution from the combined catalogue as much as possible. To achieve this, we adopt the binning method. The binning scheme is determined through the forecasting result using Fisher matrix method. We split the whole sample into $\mathit{nine}$ $\mathit{overlapping}$ redshift bins to make sure that the measurement precision of the isotropic BAO signal is better than 3\% in each bin. We perform the measurements on the an/isotropic BAO positions in the $\mathit{nine}$ $\mathit{overlapping}$ bins using the correlation functions of the pre- and post-reconstruction catalogues. To test the constraining power of our tomographic BAO measurements, we perform the fitting of cosmological parameters. 

The analysis is part of a series of papers analysing the clustering of the completed BOSS DR12 \citep{Alam2016, TomoPk, pk-BAO, pk-full, CF-sysweight, Sanchez2016, Sanchez2016-2, Albornoz2016, Mariana2016, Grieb2016,CF_4bins,Ibanez2016}. The same tomographic BAO analysis is performed using galaxy power spectrum in Fourier space \citep{TomoPk}. Another tomographic analysis is performed using the angular correlation function in many thin redshift shells and their angular cross-correlations in the companion paper, \citet{Albornoz2016}, to extract the time evolution of the clustering signal.

In Section 2, we introduce the data and mocks used in this paper. We present the forecast result in Section 3. In Section 4, we describe the methodology to measure the BAO signal using multipoles of correlation function. In Section 5, we constrain cosmological models using the BAO measurement from the post-reconstructed catalogues. Section 6 is devoted to the conclusion. In this paper, we use a fiducial $\Lambda$CDM cosmology with the parameters:  $\Omega_m=0.307, \Omega_bh^2=0.022, h=0.6777, n_s=0.96,  \sigma_8=0.8288$. The comoving sound horizon in this cosmology is $r_d^{\rm fid}=147.74 \,\rm Mpc$.

\section{Data and Mocks}
\label{sec:data}
We use the completed catalogue of BOSS DR12, which covers the redshift range from $0.2$ to $0.75$. In the North Galactic Cap (NGC),  $864,923$ galaxies over the effective coverage area of $5923.90\,\rm deg^2$ are observed and the South Galactic Cap (SGC) contains $333,081$ with the effective coverage area of $2517.65\,\rm deg^2$. The volume density distribution from observation is shown in solid curves of Figure \ref{fig:num_dis}. 

In oder to correct for observational effects, the catalogue is given a set of weights, including weights for the redshift failure, $w_{\rm zf}$, close pair due to fiber collisions, $w_{\rm cp}$ and for systematics, $w_{\rm sys}$. In addition, the FKP weight to achieve a balance between the regions of high density and low density \citep{FKP1994} is added 
\ba
w_{\rm FKP}=\frac{1}{1+n(z)P_0},
\ea
where $n(z)$ is the number density of galaxies, and $P_0$ is set to $10,000\,h^{-3}\rm Mpc^3$. Thus each galaxy is counted by adding a total weight as below
\ba
w_{\rm tot}=w_{\rm FKP}w_{\rm sys}(w_{\rm cp}+w_{\rm zf}-1).
\ea
The details about the observational systematic weights are described in \citet{CF-sysweight}.

The correlation function is measured by comparing the galaxy distribution to a randomly distributed catalogue, which is reconstructed with the same radial selection function as the real catalogue, but without clustering structure. We use a random catalogue consisting of $50$ times random galaxies of the observed sample. 

During the cosmic evolution, non-linear structure formation and redshift space distortions (RSD) can weaken the significance of the BAO peak thus degrade the precision of BAO measurements. The BAO signal can be boosted to some extent by the reconstruction procedure, which effectively moves the galaxies to the positions as if there was no RSD and nonlinear effects \citep{Eisenstein2007}. We will also present BAO measurements using the catalogue, which is reconstructed through the reconstruction algorithm as described in \citet{Padmanabhan2012}.

Mock galaxy catalogues are required to determine the data covariance matrix, and to test the methodology. We use the MultiDark-Patchy mock catalogues \citep{Kitaura2016}. The mock catalogues are constructed to match the observed data on the angular selection function, redshift distribution, and clustering statistics (\eg \,2-point and 3-point correlation functions). We utilise $2045$ mock catalogues for the pre-reconstruction, and $1000$ mocks for the post-reconstruction. We perform the measurement for each mock catalogue, then estimate the covariance matrix of data correlation function using the method proposed in \citet{Percival2014}.

\section{BAO forecasts}
\label{sec:forecasts}
We first determine the binning scheme through the Fisher matrix method. We use the Fisher matrix formulism in \citep{Tegmark1997,FisherBAO} to predict the BAO distance parameters. Starting with the galaxy power spectrum, $P(k,\mu)$, the fisher matrix is 
\ba
F_{ij}=\int_{-1}^{1} \int_{k_{\rm min}}^{k_{\rm max}}\frac{\partial \ln P(k,\mu)}{\partial p_i} \frac{\partial \ln P(k,\mu)}{\partial p_j} V_{\rm eff}(k,\mu) \frac{k^2 dk d\mu}{8\pi^2} ,
\ea
here we set $k_{\rm min}=2\pi/V_{\rm sur}^{1/3}\hompc$ and $k_{\rm max}=0.3\hompc$.

In order to ensure that the isotropic BAO measurement precision in each bin is better than 3\%, we split the whole redshift range, \ie\,$[0.2, 0.75]$ into $9$ overlapping bins. The width of the first and last bins is $0.19$, and other bins have the same bin width, \ie\,$\Delta z=0.15$.  

\begin{figure}
\centering
{\includegraphics[scale=0.23]{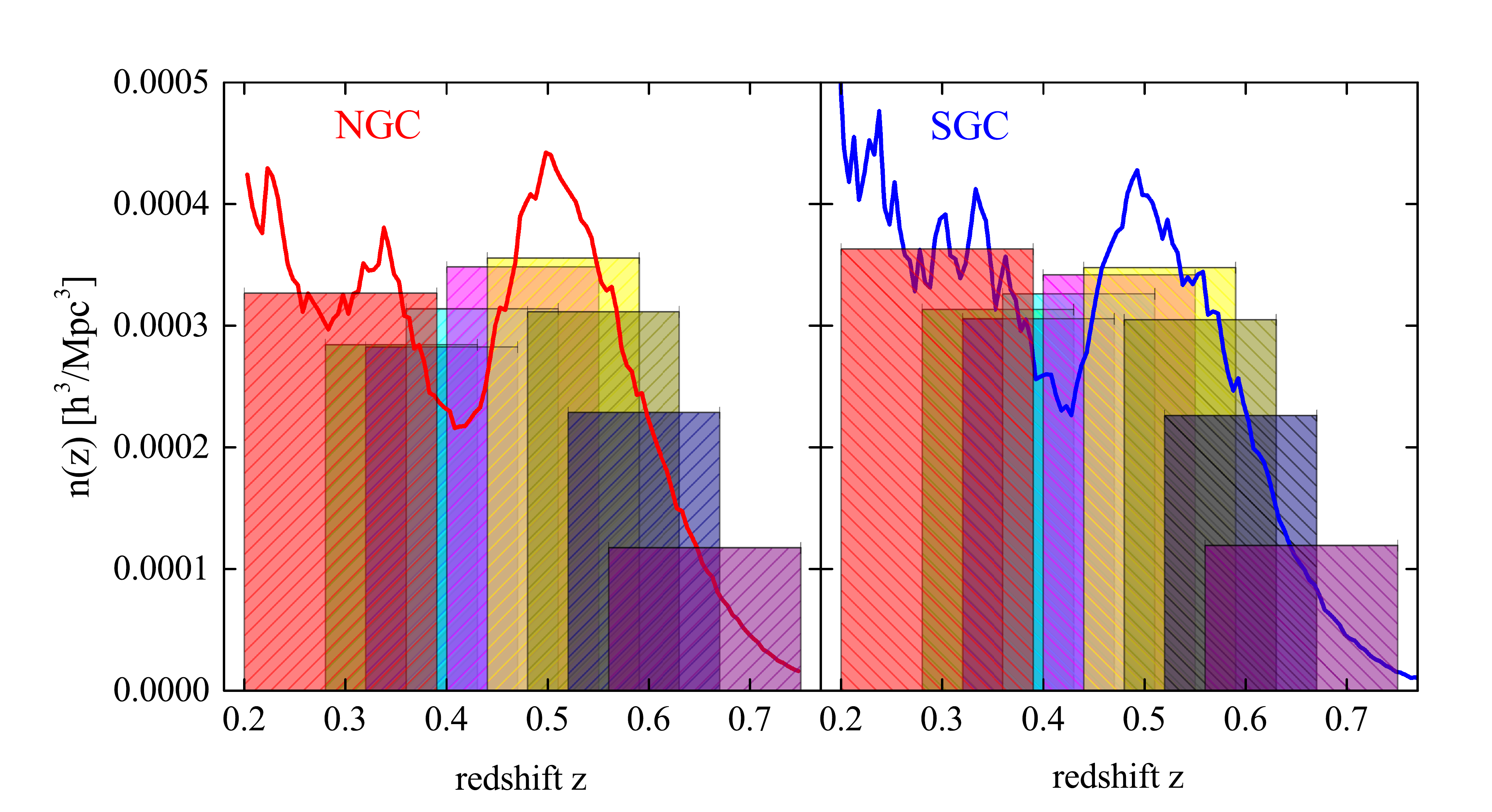}}
\caption{The overlapping histograms in different colours are the average number densities in 9 redshift bins, which is used to do the forecasts. The solid lines are the number densities for the NGC/SGC samples.}
\label{fig:num_dis}
\end{figure}

In Table \ref{tab:num}, we present the 9 overlapping redshift ranges, the effective redshifts and numbers of the samples in the NGC and SGC. In Figure \ref{fig:num_dis}, the overlapping histograms denote the average number density in each bin.

Combining the results of NGC and SGC samples as,
\ba
F^{\rm NGC+SGC}_{ij}= F^{\rm NGC}_{ij}+F^{\rm SGC}_{ij},
\ea
we present the forecast result on the precision of the BAO distance parameters, including the angular diameter distance $D_A(z)$, Hubble parameter $H(z)$ and volume distance $D_V(z)$ in Table \ref{tab:BAOforecast}. It is seen that the isotropic BAO prediction in each bin can reach, $\sigma_{D_V}/D_V < 3\%$. With the ``50\%'' reconstructed efficiency, which means that the nonlinear damping scales, $\Sigma_\perp$ and $\Sigma_{\|}$, are reduced by a factor $0.5$ and there is the remaining $50\%$ nonlinearity,  the isotropic BAO precision is within $0.8\%-1.2\%$.

\begin{figure}
\centering
{\includegraphics[scale=0.3]{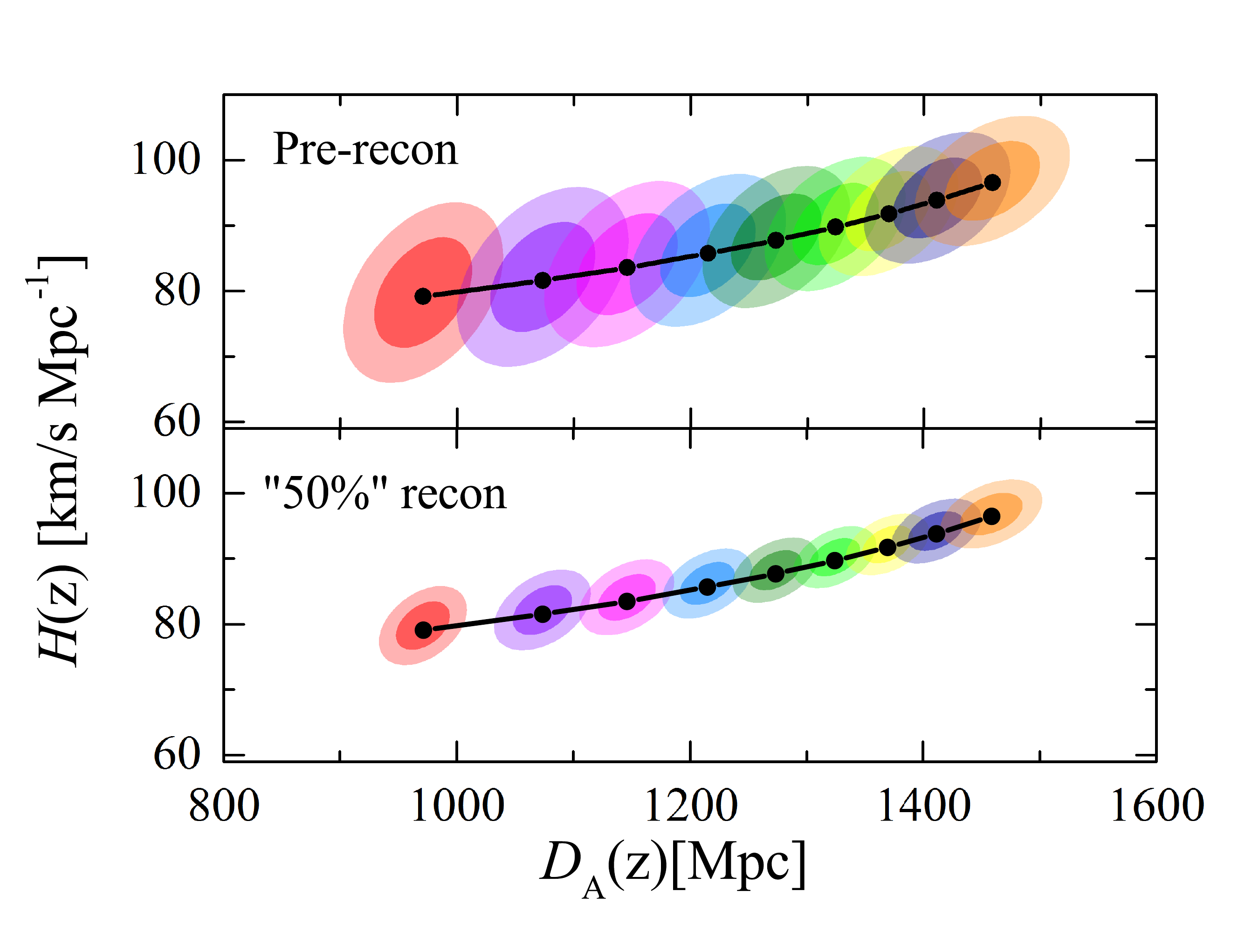}}
\caption{The 68 and 95\% CL contour plots of the transverse and radial distance parameters, $D_A(z)$ and $H(z)$, in 9 redshift bins are shown one by one from left to right. The left panel shows the result without the reconstruction, and the right panel is the result with ``50\%'' reconstructed efficiency.}
\label{fig:DA_H_forecast}
\end{figure}

The predictions on the precision of anisotropic BAO parameters are within $1.8\%-2.9\%$ for the angular diameter distance and $4.2\%-7.1\%$ for the Hubble parameter without the reconstruction. Considering the ``50\%'' reconstruction, the best prediction can reach $1.1\%$ for $D_A(z)$ and $2.1\%$ for $H(z)$. The contour plot of $D_A(z)$ and $H(z)$ within $2\,\sigma$ error is displayed in Figure \ref{fig:DA_H_forecast}, where the black points are the fiducial values. The left panel in Figure \ref{fig:DA_H_forecast} shows the forecast result without reconstruction, and the right panel presents the ``50\%'' reconstructed result.

\begin{table}
\begin{center}
\begin{tabular}{ c c c c}
\hline\hline
$z$ bins& $z_{\rm eff}$& NGC & SGC   \\ \hline
$0.20<z<0.39$& 0.31 &  208517  &  89242 \\
$0.28<z<0.43$& 0.36 &  194754  &  81539 \\
$0.32<z<0.47$& 0.40 &  230388  &  93825 \\
$0.36<z<0.51$& 0.44 &  294749  & 115029 \\
$0.40<z<0.55$& 0.48 &  370429  & 136117 \\
$0.44<z<0.59$& 0.52 &  423716  & 154486 \\
$0.48<z<0.63$& 0.56 &  410324  & 149364 \\
$0.52<z<0.67$& 0.59 &  331067  & 121145 \\
$0.56<z<0.75$& 0.64 &  243763  &  91170 \\
\hline\hline
\end{tabular}
\end{center}
\caption{The 9 overlapping redshift bins, the effective redshift and the number of samples in each bin.}
\label{tab:num}
\end{table}

\begin{table}
\caption{The forecast results on the BAO distance parameters without reconstruction (and ``50\%'' reconstruction) using the combination of NGC and SGC samples.}
\begin{center} 
\begin{tabular}{cccc}
\hline\hline
$z_{\rm eff}$  & $\sigma_{D_A}/D_A $  & $\sigma_{H}/H $  &$\sigma_{D_V}/D_V$    \\ \hline
0.31	&	0.0289 (159)	&	0.0705 (309)	&	0.0236 (114)	\\				
0.36	&	0.0281 (159)	&	0.0681 (307)	&	0.0229 (113)	\\				
0.40	&	0.0254 (145)	&	0.0616 (281)	&	0.0207 (104)	\\				
0.44	&	0.0226 (130)	&	0.0553 (253)	&	0.0185 (093)	\\				
0.48	&	0.0203 (118)	&	0.0502 (230)	&	0.0167 (085)	\\				
0.52	&	0.0188 (110)	&	0.0464 (214)	&	0.0155 (079)	\\				
0.56	&	0.0180 (108)	&	0.0441 (208)	&	0.0147 (077)	\\				
0.59	&	0.0183 (113)	&	0.0436 (214)	&	0.0147 (080)	\\				
0.64	&	0.0187 (122)	&	0.0418 (222)	&	0.0144 (085)	\\								               
\hline  \hline                       
\end{tabular}
\end{center}
\label{tab:BAOforecast}
\end{table}

\section{BAO measurements}

\subsection{The estimator for the 2-pt correlation function}
We measure the correlation function of the combined sample using the \citet{LSestimator} estimator:
\ba
\label{eq:xi_Landy_pre}
\xi(s,\mu) = \frac{DD(s,\mu)-2DR(s,\mu)+RR(s,\mu)}{RR(s,\mu)},
\ea
where DD, DR and RR are the weighted data-data pair counts, data-random pair counts and random-random pair counts with the separation, $s$ and the cosine of the angle of the pair to the line of sight, $\mu$. 

The multipole projections of the correlation function can be calculated through
\ba
\label{eq:multipole_CF}
\xi_l(s) = \frac{2l+1}{2}\int_{-1}^{1}{\rm d}\mu\, \xi(s,\mu) \mathcal{L}_{\ell}(\mu),
\ea
where $\mathcal{L}_{\ell}(\mu)$ is the Legendre Polynomial.

We also measure the correlation function of the reconstructed catalogue using  the \citet{LSestimator} estimator:
\ba
\label{eq:xi_Landy_post}
\xi(s,\mu) = \frac{DD(s,\mu)-2DS(s,\mu)+SS(s,\mu)}{RR(s,\mu)},
\ea
here we used the shifted data and randoms for $DD$, $DS$, and $SS$.

\begin{figure}
\centering
{\includegraphics[scale=0.2]{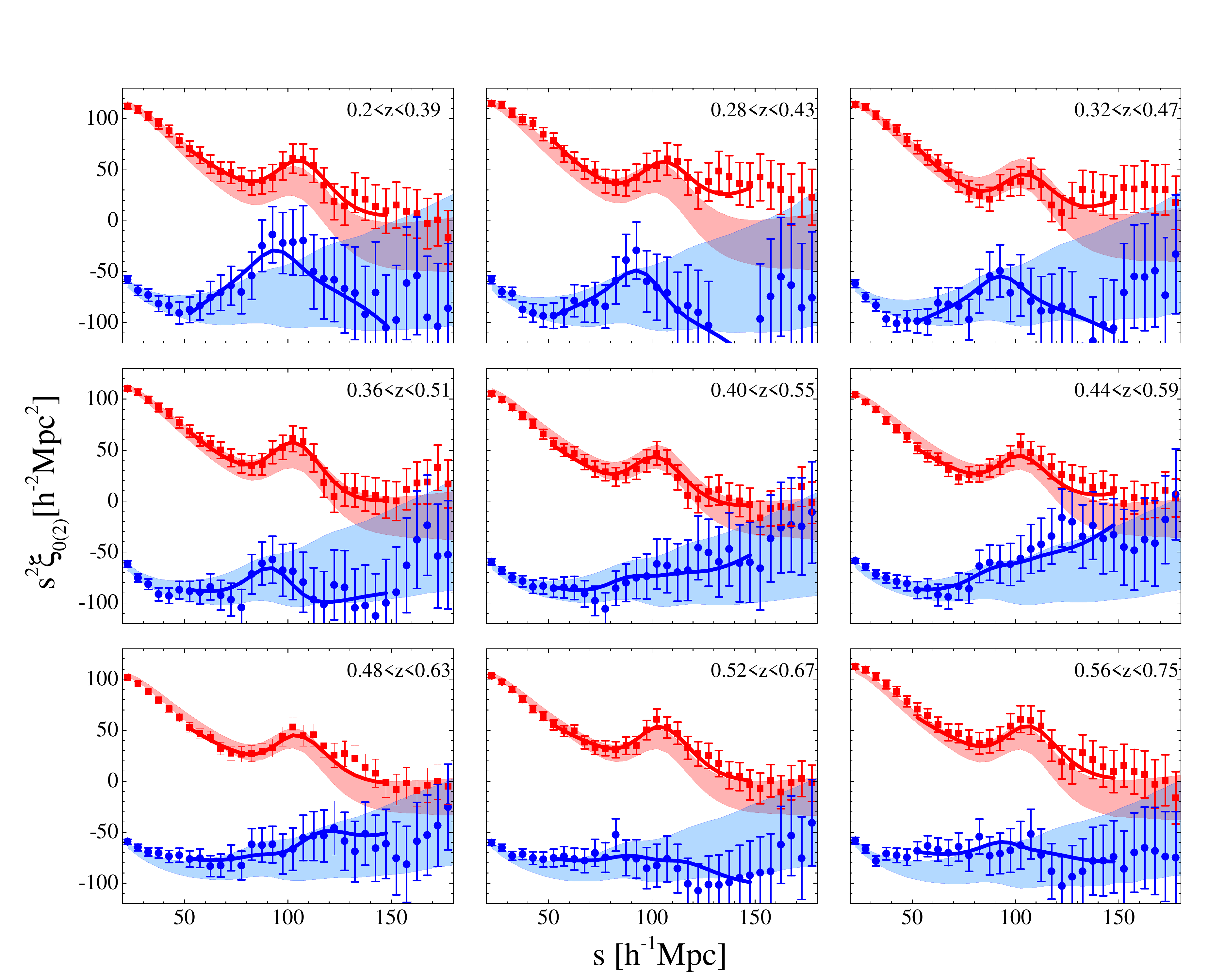}} \\
\caption{The measured monopole and quadrupole of correlation function using the pre-reconstructed catalogue in each redshift bin: in each panel the red square with $1\,\sigma$ error bar is the measured monopole and the red shaded band is the average of monopoles from mocks with a standard deviation. The blue point with $1\,\sigma$ error bar is the measured quadrupole and the blue shaded band is the average of quadruples from mocks with a standard deviation. The solid lines show the fitting results. }
\label{fig:pre_NS_bins}
\end{figure}

\begin{figure}
\centering
{\includegraphics[scale=0.2]{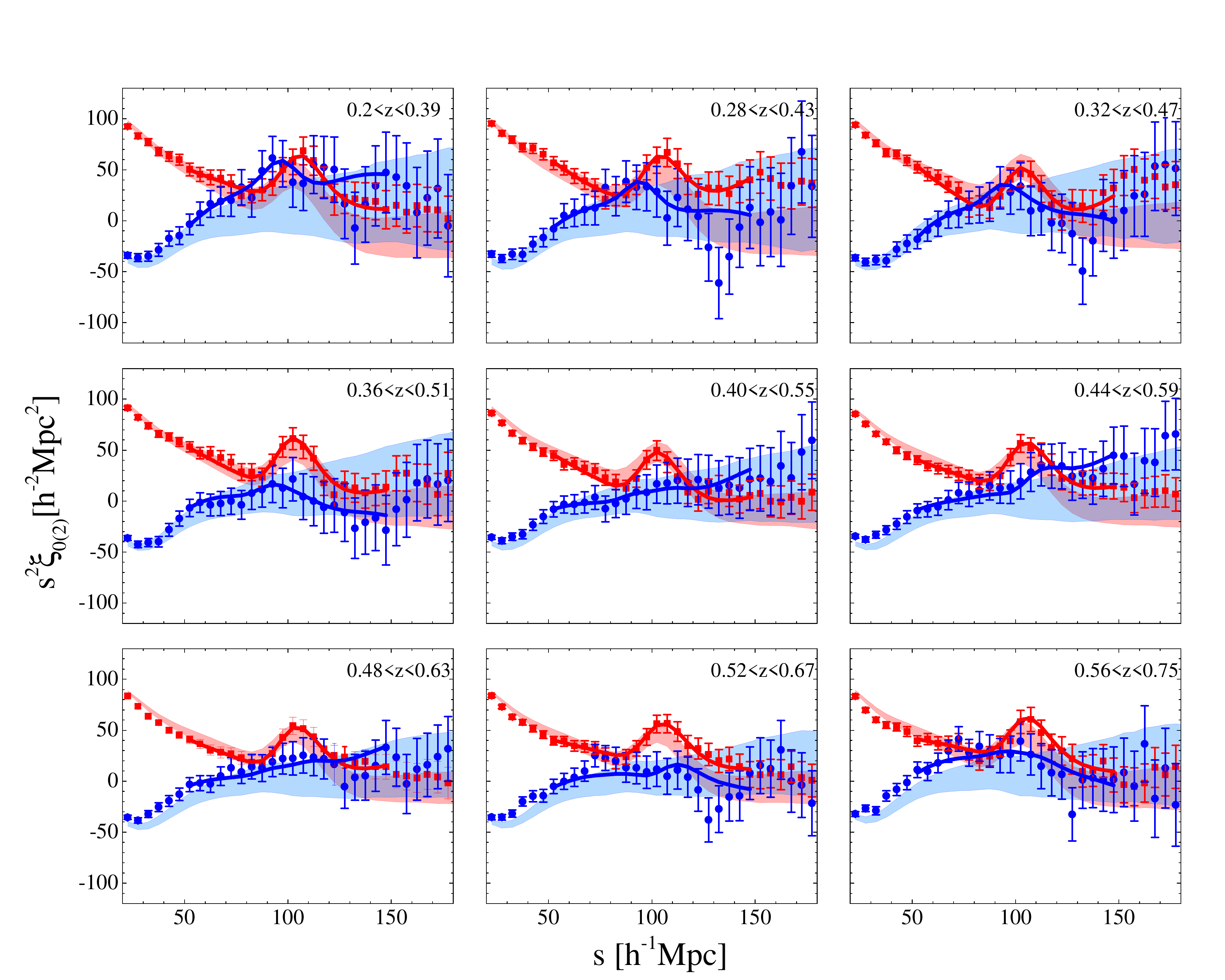}} \\
\caption{The measured monopole and quadrupole of correlation function using the post-reconstructed catalogue in each redshift bin: in each panel the red square with $1\,\sigma$ error bar is the measured monopole and the red shaded band is the average of monopoles from mocks with a standard deviation. The blue point with $1\,\sigma$ error bar is the measured quadrupole and the blue shaded band is the average of quadruples from mocks with a standard deviation. The solid lines show the fitting results. }
\label{fig:post_NS_bins}
\end{figure}

The measured monopole and quadruple of correlation function from data and mocks in each redshift bin are shown in Figure \ref{fig:pre_NS_bins} for the pre-reconstruction measurements, and in Figure \ref{fig:post_NS_bins} for the post-reconstruction measurements, where the red squares with $1\,\sigma$ error bar are the measurements of monopole from data. The red shaded regions correspond to the standard deviation from the mocks around the average. The blue points with $1\,\sigma$ error bar are the data measurements of quadrupole, and the blue shaded regions denote the average with a standard deviation from the mocks. 

\begin{figure*}
\centering
{\includegraphics[scale=0.23]{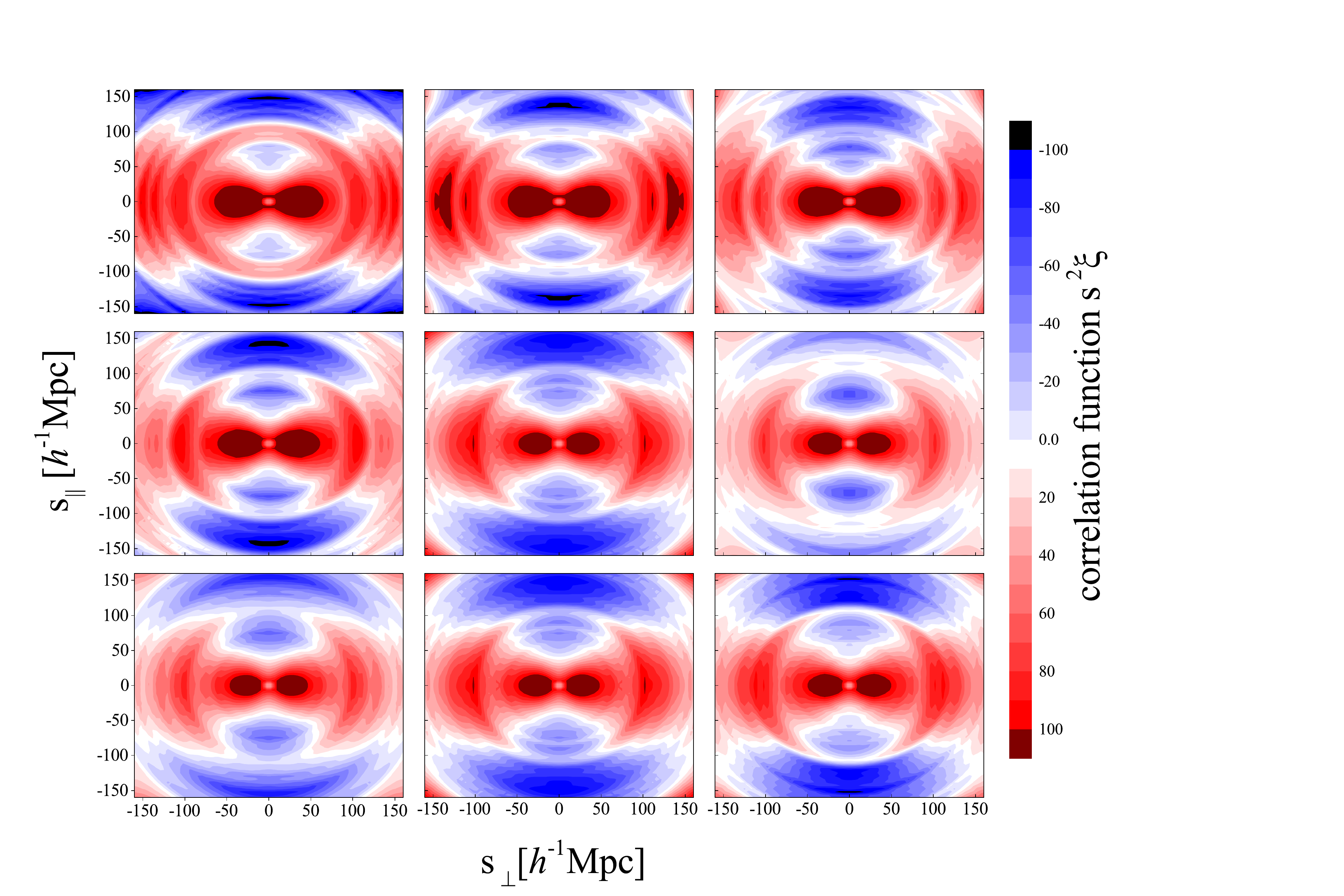}} {\includegraphics[scale=0.23]{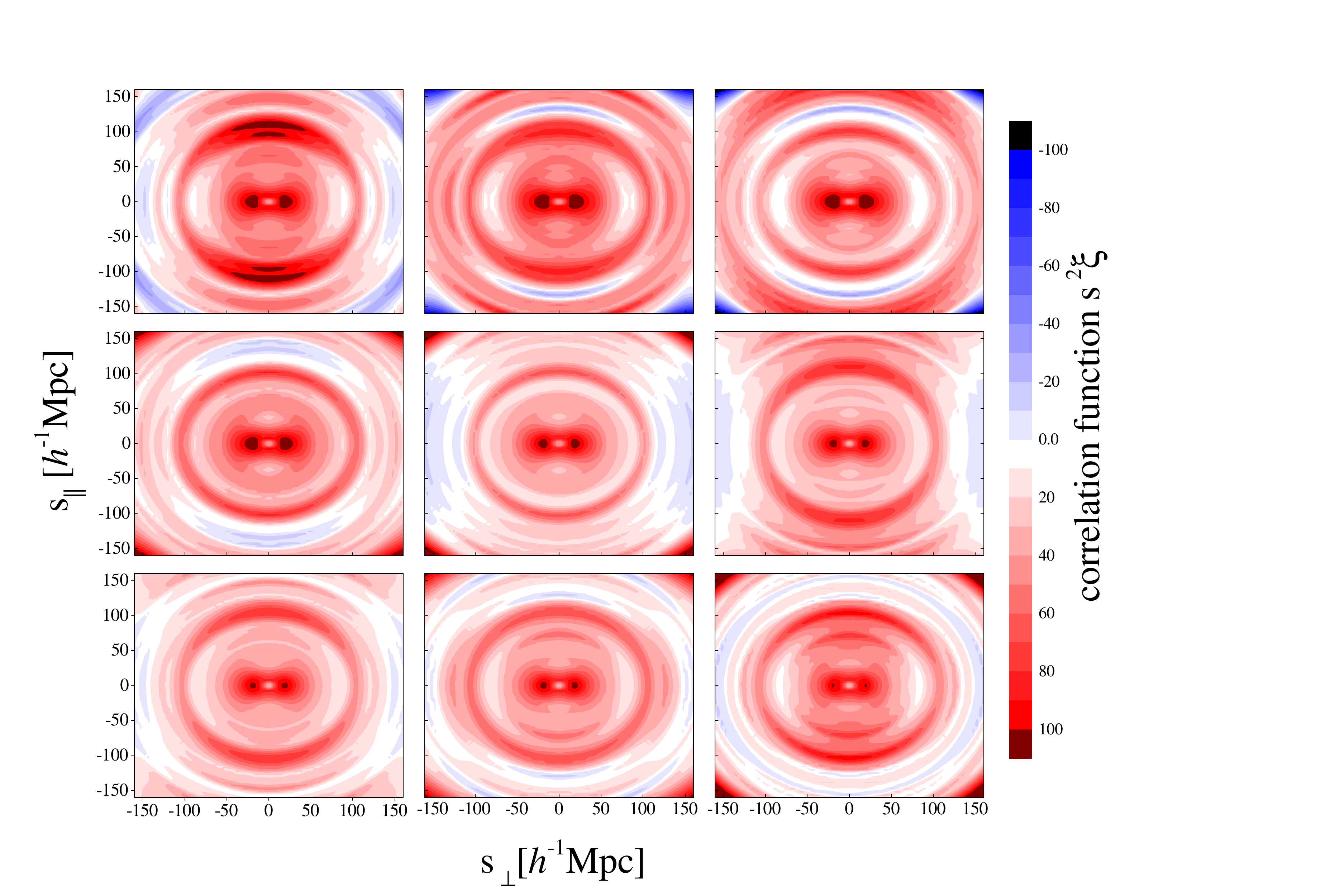}} \\
\caption{The 2D pre-reconstruction correlation functions (left panel) and  post-reconstruction correlation functions (right panel) in 9 redshift bins, which is assembled using the measured monopole and quadruple from the NGC and SGC samples, i.e. $\xi(s,\mu) = \xi_0(s) \mathcal{L}_0(\mu)+ \xi_2(s) \mathcal{L}_2(\mu)$, here $s_{\|} =s \mu$ and $s_{\perp} = s \sqrt{1-\mu^2}$.}
\label{fig:pre-xi2D}
\end{figure*}

The 2D correlation functions measured in 9 redshift bins using the pre-reconstructed and post-reconstructed catalogues are plotted in Figure \ref{fig:pre-xi2D}, where the BAO ring in each redshift slice is visualised. As expected, the BAO ring becomes clear after reconstruction. 

\subsection{The template}
The isotropic BAO position is parameterised by the scale dilation parameter,
\ba
  \alpha \equiv \frac{D_V(z)r_{d,{\rm fid}}}{D^{\rm fid}_V(z) r_d} \,.
\ea

We adopt the template for the correlation function in the isotropic case \citep{Eisenstein07},
\ba \label{eq:xi_mod}
 \xi^{\rm mod}(s) = \int \frac{k^2dk}{2\pi^2}P_{\rm dw}^{\rm mod}(k)F(k,\Sigma_s) j_0(ks) \,,
\ea
where the damping term is given by
\ba
F(k, \Sigma_s) = \frac{1}{(1+k^2\Sigma_s^2/2)^2} \,.
\ea
Here we set the parameter $\Sigma_s=4 \mpcoh$, which is the same as used in \citep{CF-sysweight}. The de-wiggled power spectrum, $P_{\rm dw}^{\rm mod}(k)$, is given by
\ba \label{eq:pk_mod}
  P_{\rm dw}^{\rm mod}(k) = P^{\rm nw}(k)+\left[P^{\rm lin}(k)-P^{\rm nw}(k)\right] e^{-\frac{1}{2}k^2 \Sigma_{\rm nl}^2}\,,
\ea
where $P^{\rm nw}(k)$ is the ``no-wiggle'' power spectrum, where the BAO feature is erased,  which is obtained using the fitting formulae in \citet{Eisenstein}. The linear power spectrum $P^{\rm lin}(k)$ is calculated by CAMB\footnote{\url{http://camb.info}} \citep{CAMB}. $\Sigma_{\rm nl}$ in the Gaussian term is a damping parameter.  

Then, allowing an unknown bias factor $B_\xi$, which rescales the amplitude of the input template, the correlation function is given by 
\ba \label{eq:xi_iso_fit}
\xi^{\rm fit}(s) = B_\xi^2\xi^{\rm mod}(\alpha s)+A^\xi(s)\,,
\ea
which includes the polynomial terms for systematics
\ba
 A^\xi(s) = \frac{a_1}{s^2} + \frac{a_2}{s} + a_3\,.
\ea
Before doing the fitting, we normalise the model to the data at the scale $s=50 \mpcoh$, as done in \citet{Xu2013, Anderson2014} . While performing the fitting, we add a Gaussian prior on $\rm log(B_\xi^2)=0\pm0.4$ \citep{Xu2013,Anderson2014}. So in the isotropic case, we have 5 free parameter, i.e. \,[log$(B_\xi^2),\alpha,a_1,a_2,a_3$].

The BAO feature can be measured in both the transverse and line-of-sight directions. This can be parametrised by $\alpha_\perp$ and $\alpha_{||}$, respectively
\ba
\alpha_{\perp} = \frac{D_{A}(z)r_d^{\rm fid}}{D^{\rm fid}_{A}(z) r_d} \,,\,\,\,\,
\alpha_{\parallel} = \frac{H^{\rm fid}(z)r_d^{\rm fid}}{H(z) r_d} \,.
\ea

The anisotropic correlation function is modelled as a transform of the 2D power spectrum, 
\ba
P(k,\mu) = (1+\beta\mu^2)^2 F(k,\mu,\Sigma_s)P_{\rm dw}(k,\mu),
\ea
where the $(1+\beta\mu^2)^2$ term corresponds to the Kaiser model for large-scale RSD \citep{Kaiser}. For the reconstruction, this term is replaced by $[1+\beta\mu^2(1-S(k))]^2$ with the smoothing, $S(k)=e^{-k^2\Sigma_r^2/2}$ and $\Sigma_r=15 \mpcoh$ \citep{Seo2015}. The term
\ba
F(k,\mu,\Sigma_s) = \frac{1}{(1+k^2\mu^2\Sigma_s^2/2)^2}
\ea
is introduced to model the small-scale FoG effect. The 2D de-wiggled power spectrum, compared to Eq. \ref{eq:pk_mod}, becomes
 \ba
 P_{\rm dw}(k,\mu) &=& [P_{\rm lin}(k) - P_{\rm nw}(k)] \nonumber \\
&& \cdot \exp \bigg[
-\frac{k^2\mu^2\Sigma_\parallel^2 +k^2(1-\mu^2)\Sigma_\perp^2}{2} \bigg]
+P_{\rm nw}(k), \nonumber \\
 \ea
here the Gaussian damping term is also anisotropic. $\Sigma_\parallel$ and $\Sigma_\perp$ are the line-of-sight
and transverse components of $\Sigma_{\rm nl}$, \ie\,$\Sigma_{\rm nl}^2
= (\Sigma_\parallel^2 + 2\Sigma_\perp^2)/3$. Here we set  $\Sigma_\parallel=4 \mpcoh$ and $\Sigma_\perp=2.5 \mpcoh$ for the post-reconstruction and  $\Sigma_\parallel=10 \mpcoh$ and $\Sigma_\perp=6 \mpcoh$ for the pre-reconstruction \citep{CF-sysweight}. 

Given the 2D power spectrum $P(k,\mu)$, which can be decomposed into Legendre moments, then the multipoles of power spectrum are 
\ba
P_{\ell}(k) = \frac{2\ell+1}{2}\int^1_{-1} P(k,\mu) \mathcal{L}_\ell(\mu) d\mu,
\ea
which can be transformed to the multipoles of correlation function by
\ba
\xi_{\ell}(s) = \frac{i^{\ell}}{2\pi^2} \int k^2 P_{\ell}(k) j_\ell(ks) dk.
\ea
Using the Legendre polynomials, we have 
\ba
\xi(s,\mu) = \sum_{\ell} \xi_{\ell}(s) \mathcal{L}_\ell(\mu). 
\ea

Then the model multipoles of correlation function are
\ba
\xi_{\ell}(s,\alpha_{\perp},\alpha_{\parallel})= \frac{2\ell+1}{2}\int^1_{-1} \xi(s',\mu') \mathcal{L}_\ell(\mu) d\mu,
\ea
where $s'=s \sqrt{\mu^2\alpha^2_{\parallel}+(1-\mu^2)\alpha^2_{\perp}}$ and $\mu'=\mu \alpha_{\parallel}/\sqrt{\mu^2\alpha^2_{\parallel}+(1-\mu^2)\alpha^2_{\perp}}$ are respectively the separation between two galaxies and the cosine of the angle of the pair to the line of sight in the true cosmology, .

In addition, we use a bias parameter $B_0$ to adjust the amplitude of the input template and include the model for systematics using the polynomial terms
\ba
A_\ell(s) = \frac{a_{\ell,1}}{s^2} + \frac{a_{\ell,2}}{s} + a_{\ell,3}.
\ea
So we fit the data using the model multipoles
\ba
\xi^{\rm mod}_0(s)&=&B_0\xi_0(s,\alpha_{\perp},\alpha_{\parallel})+A_0(s), \\
\xi^{\rm mod}_2(s)&=&\xi_2(s,\alpha_{\perp},\alpha_{\parallel})+A_2(s).
\ea
As in the isotropic case, the monopole template is normalised to the measurement at $s=50 \mpcoh$. So in the anisotropic case, we have 10 free parameter, \ie\,$[\alpha_{\perp}, \alpha_{||}$, log$(B_0^2), \beta, a_{\ell,1-3}]$. While performing the fitting, a Gaussian prior on log$(B_0^2)=0\pm0.4$ is applied. We also add a Gaussian prior for the RSD parameter, \ie\, $\beta=0.4\pm0.2$ \citep{Anderson2014}. 

\subsection{Covariance matrix}
When fitting the BAO parameters, $\bold{p}$, we use the MCMC to search for the minimum $\chi^2$,
\ba
 \chi^2 (\bold{p}) \equiv  \sum_{i,j}^{\ell,\ell'}  \left[\xi^{th}_{\ell} (s_i, \bold{p}) -\xi_{\ell}(s_i) \right] F^{\ell,\ell'}_{ij} \left[\xi_{\ell'}^{th}(s_j, \bold{p}) -\xi_{\ell'}(s_j)\right]. \nonumber
\ea
where $F^{\ell,\ell'}_{ij}$ is the inverse of the covariance matrix, $C^{\ell,\ell'}_{ij}$, which is estimated using mock catalogues,
\ba
C^{\ell,\ell'}_{ij}=\frac{1}{N-1} \sum_{k} \left[\xi_{\ell}^k(s_i)-\bar{\xi}_{\ell}(s_i)\right] \left[\xi_{\ell'}^k(s_j)-\bar{\xi}_{\ell'}(s_j)\right],
\ea
where the average multipoles is given by
\ba
\bar{\xi}_{\ell}(s_i)=\frac{1}{N} \sum_{k}\xi_{\ell}^k(s_i),
\ea
here $N$ is the number of mocks: $N=2045$ in the pre-reconstruction case and $N=1000$ in the post-reconstruction case. The unbiased estimation for the inverse covariance matrix is given by
\ba \label{eq:inv_cov}
\widetilde{C}_{ij}^{-1}=\frac{N-N_b-2}{N-1}C_{ij}^{-1}.
\ea
where $N_b$ is the number of the scale bins. In order to include the error propagation from the error in the covariance matrix into the fitting parameters \citep{Percival2014} we rescale the covariance matrix, $\widetilde{C}_{ij}$, by 
\ba
M=\sqrt{\frac{1+B(N_b-N_p)}{1+A+B(N_p+1)}}
\ea
here $N_p$ is the number of the fitting parameters, and 
\ba
A=\frac{2}{(N-N_b-1)(N-N_b-4)}, \\
B=\frac{N-N_b-2}{(N-N_b-1)(N-N_b-4)}.
\ea

\begin{figure}
\centering
{\includegraphics[scale=0.35]{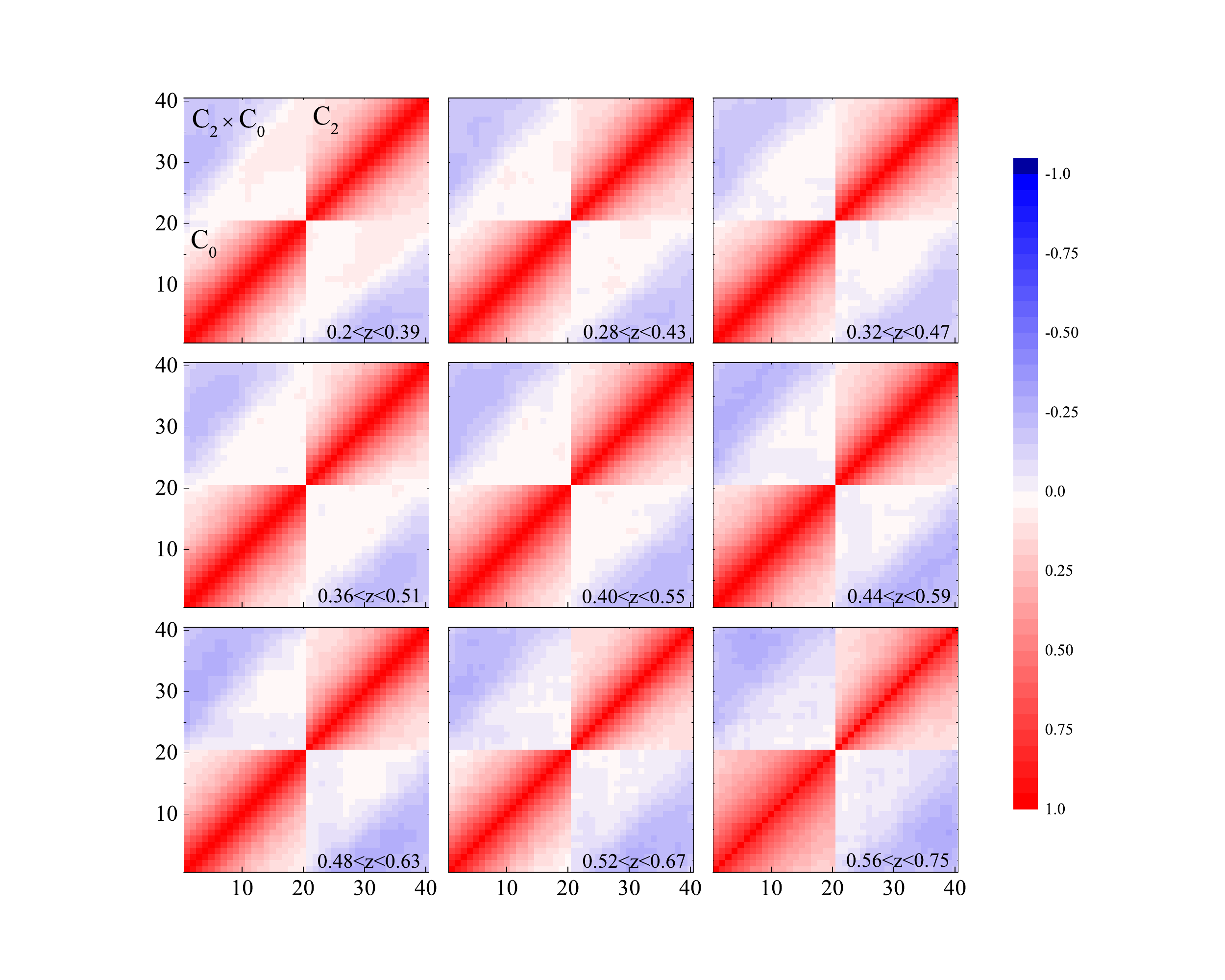}}
\caption{The correlations between monopole, $\rm C_0$, quadrupole, $\rm C_2$ and their cross correlation, $\rm C_0\times C_2$ for the pre-reconstruction. }
\label{fig:pre_cov_xi0_xi2}
\end{figure}

\begin{figure}
\centering
{\includegraphics[scale=0.35]{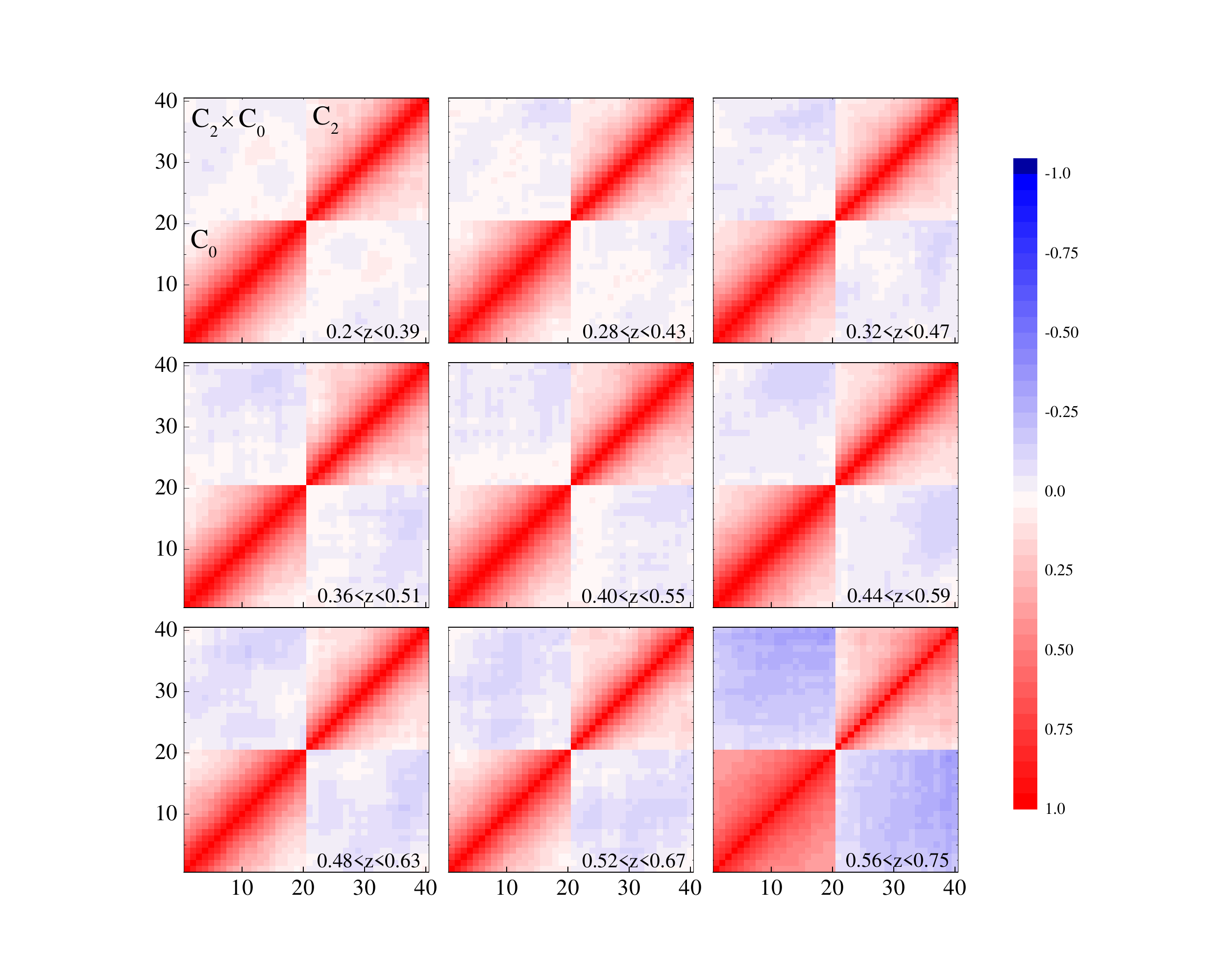}}
\caption{The correlations between monopole, $\rm C_0$, quadrupole, $\rm C_2$ and their cross correlation, $\rm C_0\times C_2$ for the post-reconstruction. }
\label{fig:post_cov_xi0_xi2}
\end{figure}

The normalised covariance matrix showing the correlations between monopole, quadrupole and their cross correlation in each bin is plotted in Figure \ref{fig:pre_cov_xi0_xi2} in the pre-reconstruction case and Figure \ref{fig:post_cov_xi0_xi2} in the post-reconstruction case. From Figure \ref{fig:pre_cov_xi0_xi2} and Figure \ref{fig:post_cov_xi0_xi2}, it is seen that after reconstruction, there is less auto-correlation of multipoles and cross-correlation between multipoles.

\section{Tests on mock catalogues}
We present the mock tests for the BAO analysis using 1000 pre-reconstructed and post-reconstructed mocks. We perform the isotropic and anisotropic BAO measurements using each individual mock catalogue in both cases. The results are shown in Table \ref{tab:pre_post_test}, where we list the average of fitting value from each mock, standard derivations, and the average of $1\sigma$ error for the parameters, $\alpha$, $\alpha_{\perp}$ and $\alpha_{\parallel}$. The fiducial cosmology we use here corresponds to the input cosmology of the mocks, therefore we expect that the average values of parameters $\alpha$, $\alpha_{\perp}$ and $\alpha_{\parallel}$ are equal to $1$. 

\begin{table*}
\caption{The statistics of the isotropic and anisotropic BAO fittings using the pre-reconstructed and post-reconstructed mocks. $\langle \alpha \rangle$, $\langle \alpha_{\perp} \rangle$ and $\langle \alpha_{\parallel} \rangle$ are the average of the fitting mean value from each mock. $S_{\alpha}$, $S_{\alpha_{\perp}}$ and $S_{\alpha_{\parallel}}$ are the standard derivation of the parameters $\alpha$, $\alpha_{\perp}$ and $\alpha_{\parallel}$, respectively. $\langle \sigma_{\alpha}\rangle$,  $\langle \sigma_{\alpha_{\perp}} \rangle$ and $\langle \sigma_{\alpha_{\parallel}}\rangle$ correspond to the average of $1\sigma$ error of these three parameters from each mock.}
\begin{center} 
\begin{tabular}{cccccccccccc}
\hline\hline
$z_{\rm eff}$  & $\langle \alpha \rangle$  & $S_{\alpha}$ &  $\langle \sigma_{\alpha}\rangle$ &$\langle \chi^2\rangle/\rm dof$& $\langle \alpha_{\perp} \rangle$  & $S_{\alpha_{\perp}}$ &  $\langle \sigma_{\alpha_{\perp}} \rangle$ & $\langle \alpha_{\parallel} \rangle$  & $S_{\alpha_{\parallel}}$ &  $\langle \sigma_{\alpha_{\parallel}}\rangle$ &$\langle \chi^2\rangle/\rm dof$   \\ \hline
\rm pre-reconstruction: & & & & & & & & & \\
$0.31$	&	0.996	&	0.033	&	0.036	&	15.1/15	&	0.997	&	0.042	&	0.044	&	0.992	&	0.064	&	0.078	&	30.3/30	\\
$0.36$	&	0.997	&	0.031	&	0.034	&	15.1/15	&	0.995	&	0.040	&	0.043	&	0.995	&	0.064	&	0.077	&	30.3/30	\\
$0.40$	&	1.000	&	0.029	&	0.031	&	15.0/15	&	0.998	&	0.038	&	0.039	&	0.996	&	0.064	&	0.073	&	30.2/30	\\
$0.44$	&	1.001	&	0.024	&	0.027	&	15.2/15	&	0.999	&	0.033	&	0.034	&	0.998	&	0.061	&	0.068	&	30.3/30	\\
$0.48$	&	1.003	&	0.022	&	0.024	&	15.2/15	&	0.999	&	0.030	&	0.030	&	1.003	&	0.061	&	0.062	&	30.2/30	\\
$0.52$	&	1.002	&	0.021	&	0.022	&	15.2/15	&	0.999	&	0.028	&	0.029	&	1.001	&	0.059	&	0.060	&	30.1/30	\\
$0.56$	&	1.002	&	0.020	&	0.022	&	15.1/15	&	0.998	&	0.029	&	0.029	&	1.003	&	0.058	&	0.059	&	29.9/30	\\
$0.59$	&	1.001	&	0.021	&	0.023	&	15.4/15	&	0.998	&	0.031	&	0.031	&	1.001	&	0.059	&	0.061	&	30.3/30	\\
$0.64$	&	1.002	&	0.022	&	0.025	&	15.4/15	&	0.999	&	0.033	&	0.034	&	1.000	&	0.059	&	0.062	&	30.5/30	\\     \hline          
\rm post-reconstruction: & & & & & & & & & \\  
$0.31$	&	0.999	&	0.019	&	0.021	&	15.2/15	&	0.993	&	0.028	&	0.027	&	0.998	&	0.049	&	0.050	&	29.7/30	\\
$0.36$	&	0.999	&	0.018	&	0.021	&	15.2/15	&	0.992	&	0.028	&	0.026	&	0.998	&	0.050	&	0.049	&	29.7/30	\\
$0.40$	&	0.999	&	0.017	&	0.019	&	15.2/15	&	0.994	&	0.026	&	0.025	&	0.999	&	0.048	&	0.045	&	29.8/30	\\
$0.44$	&	0.999	&	0.015	&	0.016	&	15.2/15	&	0.994	&	0.022	&	0.021	&	1.001	&	0.040	&	0.038	&	30.0/30	\\
$0.48$	&	1.001	&	0.013	&	0.015	&	15.2/15	&	0.995	&	0.019	&	0.018	&	1.003	&	0.036	&	0.035	&	30.1/30	\\
$0.52$	&	1.001	&	0.013	&	0.014	&	15.3/15	&	0.996	&	0.017	&	0.017	&	1.005	&	0.034	&	0.032	&	30.1/30	\\
$0.56$	&	1.002	&	0.012	&	0.013	&	15.3/15	&	0.995	&	0.018	&	0.017	&	1.006	&	0.034	&	0.032	&	30.1/30	\\
$0.59$	&	1.001	&	0.013	&	0.014	&	15.3/15	&	0.996	&	0.019	&	0.019	&	1.003	&	0.037	&	0.035	&	29.9/30	\\
$0.64$	&	1.001	&	0.015	&	0.017	&	15.2/15	&	0.995	&	0.022	&	0.022	&	1.004	&	0.040	&	0.040	&	29.9/30	\\    \hline 
 \hline                      
\end{tabular}
\end{center}
\label{tab:pre_post_test}
\end{table*}

\begin{figure}
\centering
{\includegraphics[scale=0.3]{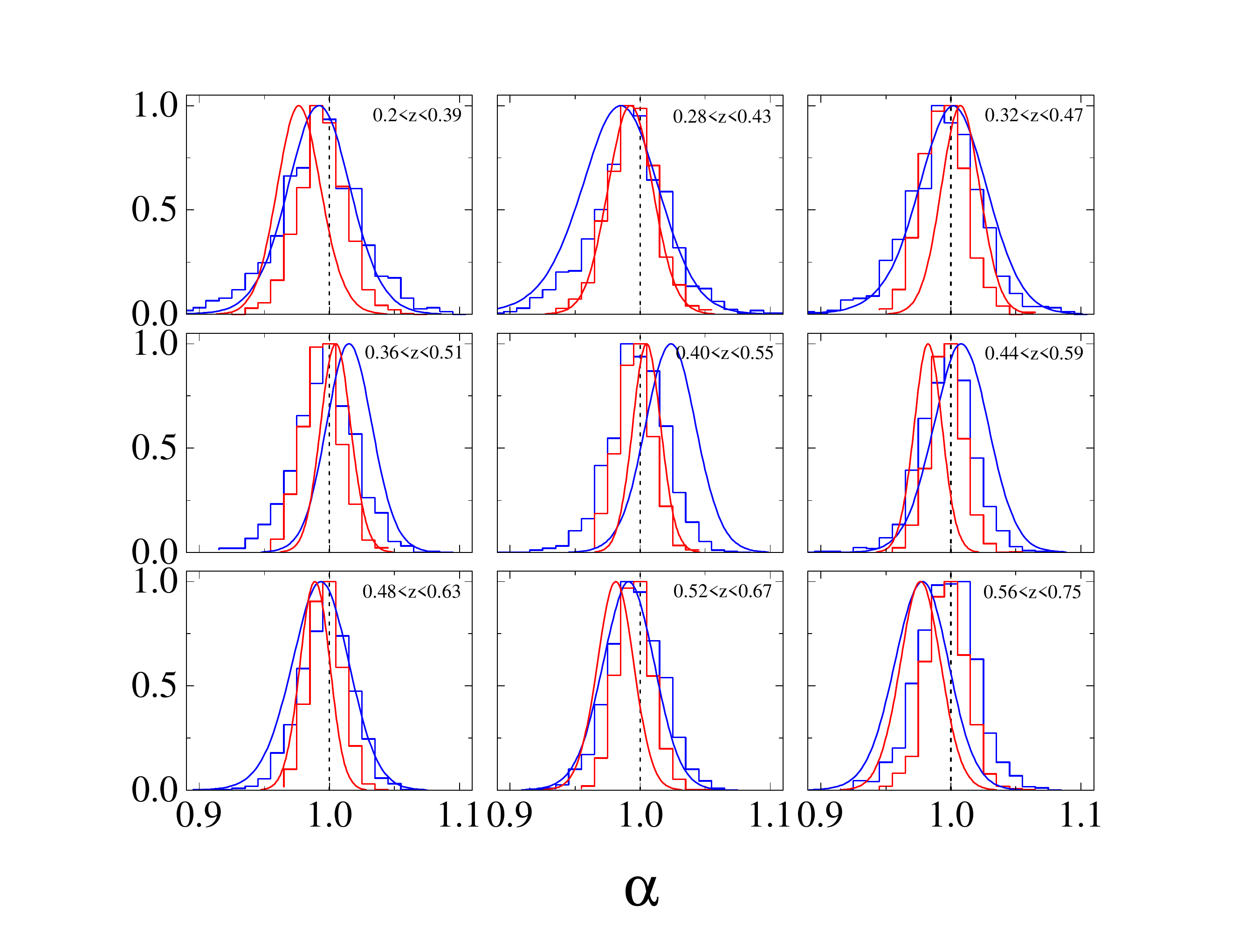}}
\caption{The 1D distribution of the parameter $\alpha$ from the pre-reconstructed mock catalogue (blue histograms), galaxy catalogue (blue curves) and from the post-reconstructed mock catalogue (red histograms), galaxy catalogue (red curves).}
\label{fig:iso_1D}
\end{figure}

\begin{figure}
\centering
{\includegraphics[scale=0.2]{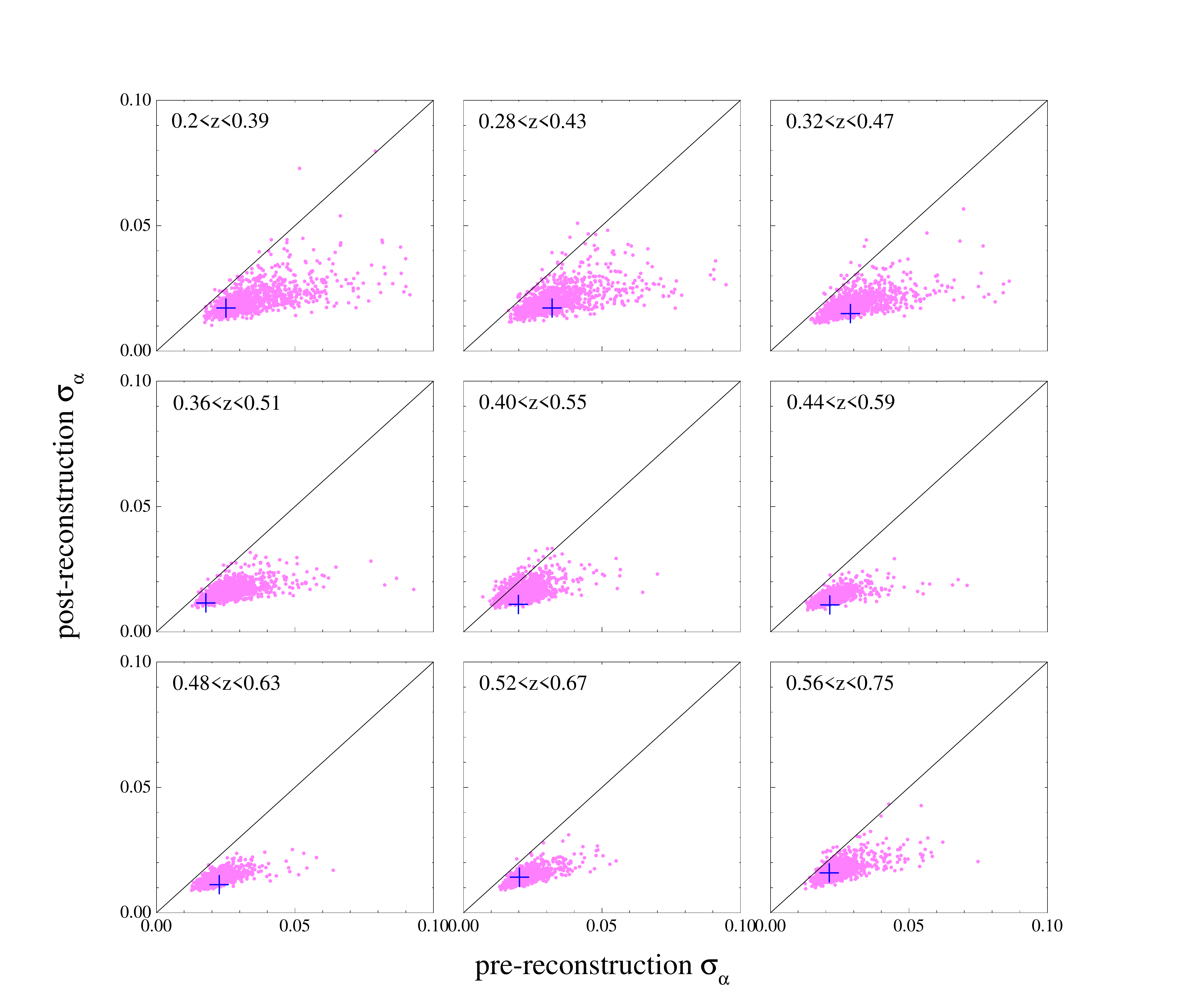}}
\caption{The scatter plot of error of $\alpha$ using pre- and post reconstruction mock catalogue. Each magenta point denotes the $1\,\sigma$ error of $\alpha$ from each mock (totally 1000 mocks) and the cross (blue) is the error measured by data. }
\label{fig:iso_mocktest}
\end{figure}

\begin{figure*}
\centering
{\includegraphics[scale=0.35]{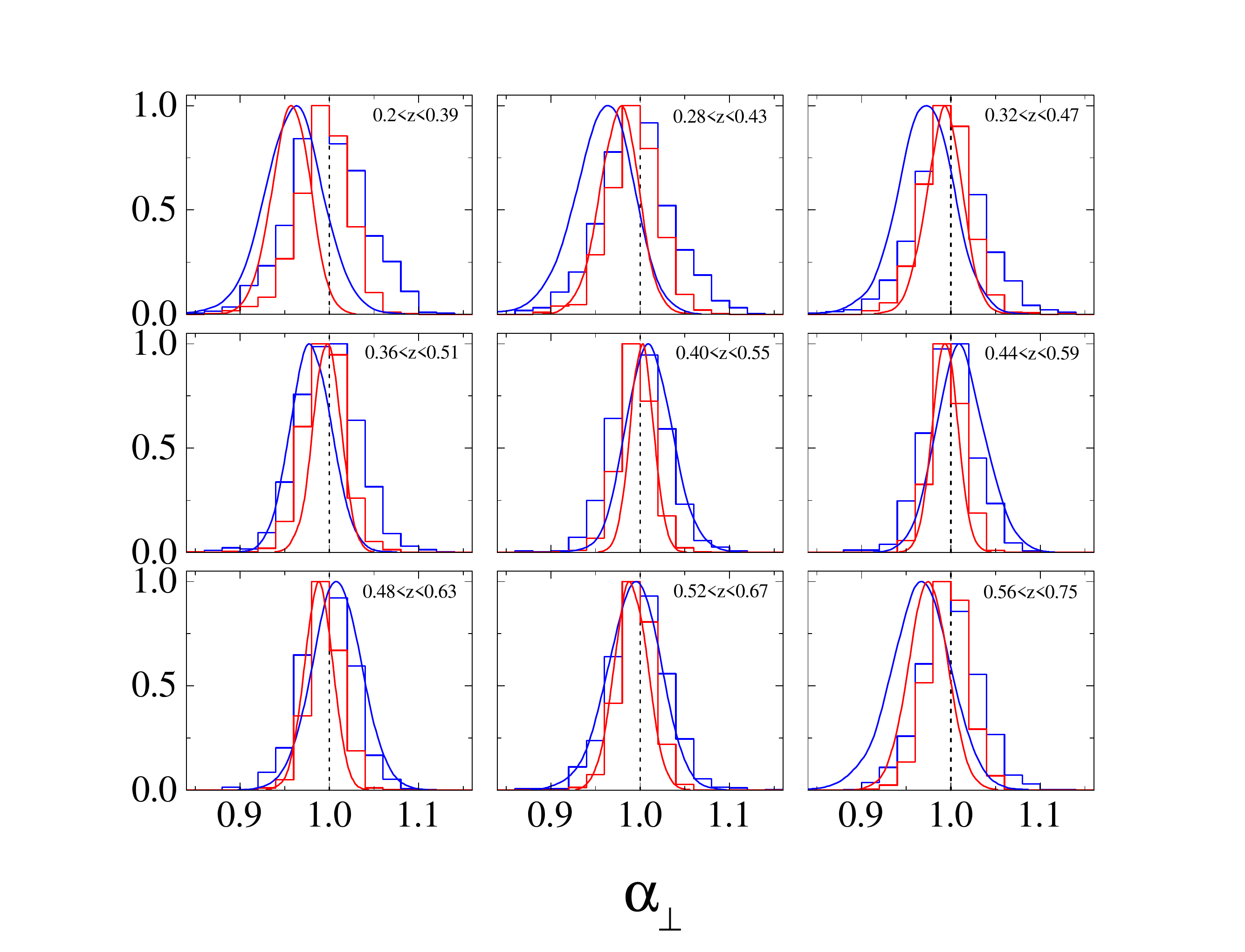}}{\includegraphics[scale=0.35]{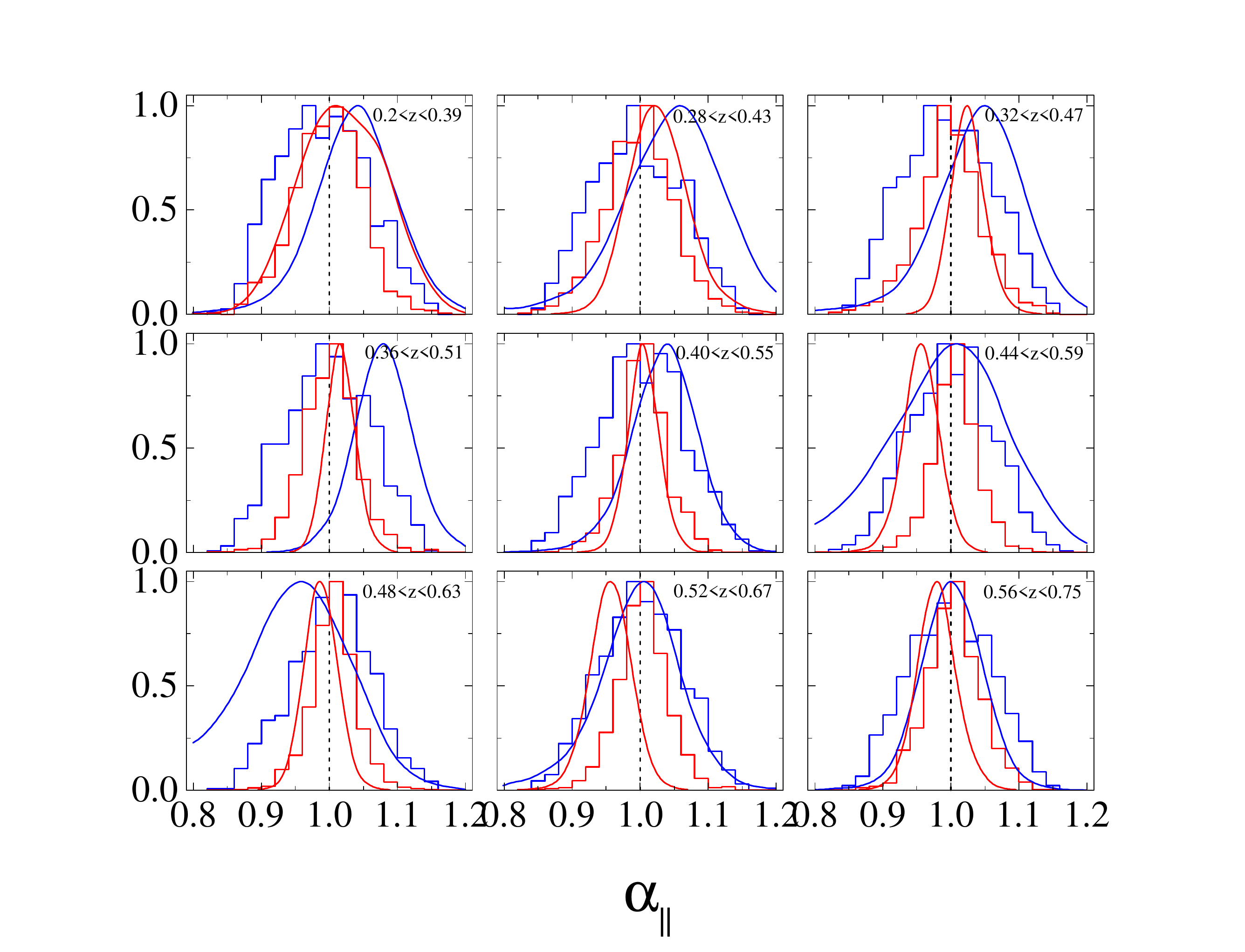}}
\caption{The 1D distributions of the parameters $\alpha_{\perp}$ (left panel) and $\alpha_{||}$ (right panel) from the pre-reconstructed mock catalogue (blue histograms), galaxy catalogue (blue curves) and from the post-reconstructed mock catalogue (red histograms), galaxy catalogue (red curves).}
\label{fig:aniso_1D}
\end{figure*}

\begin{figure*}
\centering
{\includegraphics[scale=0.2]{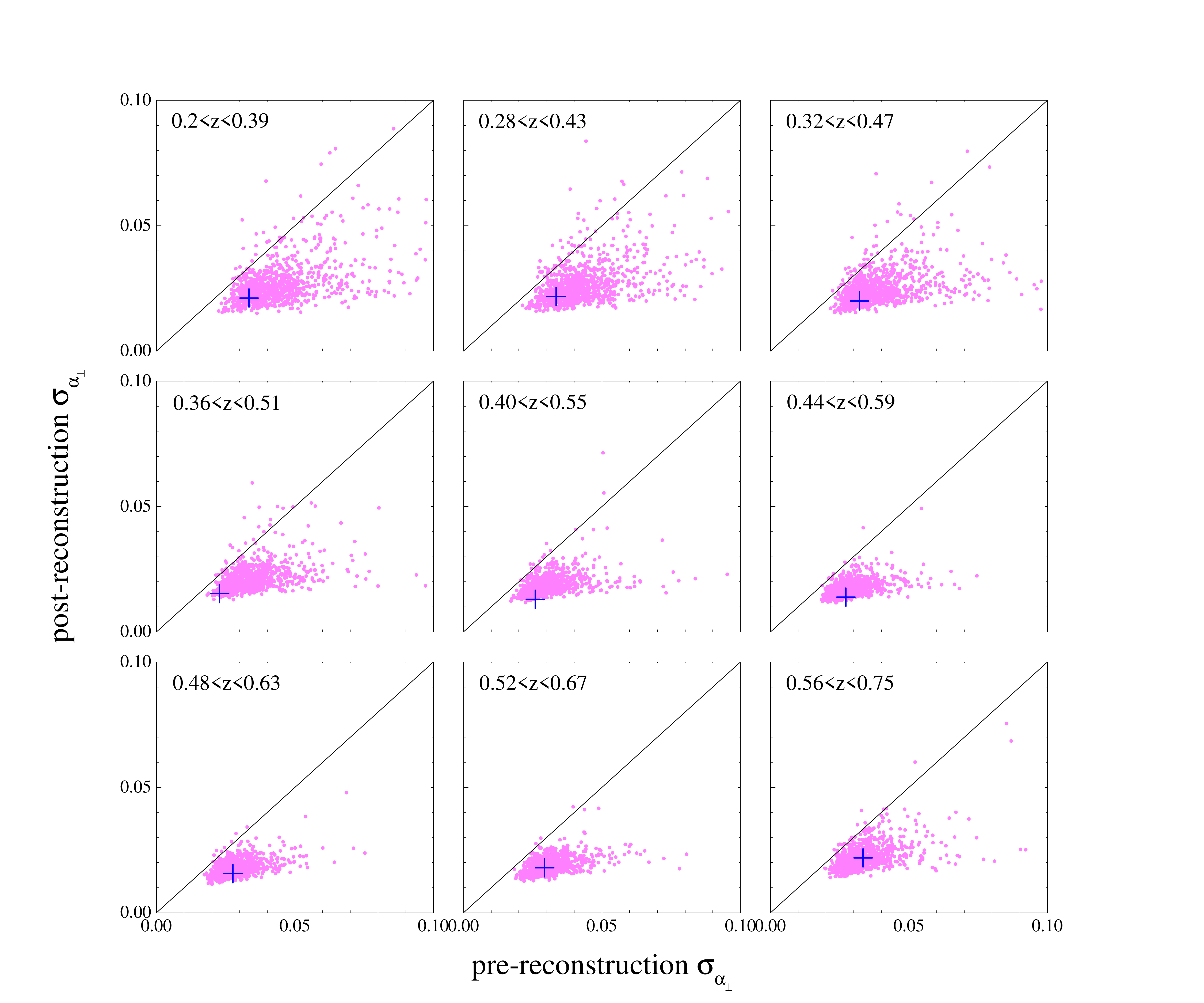}}{\includegraphics[scale=0.2]{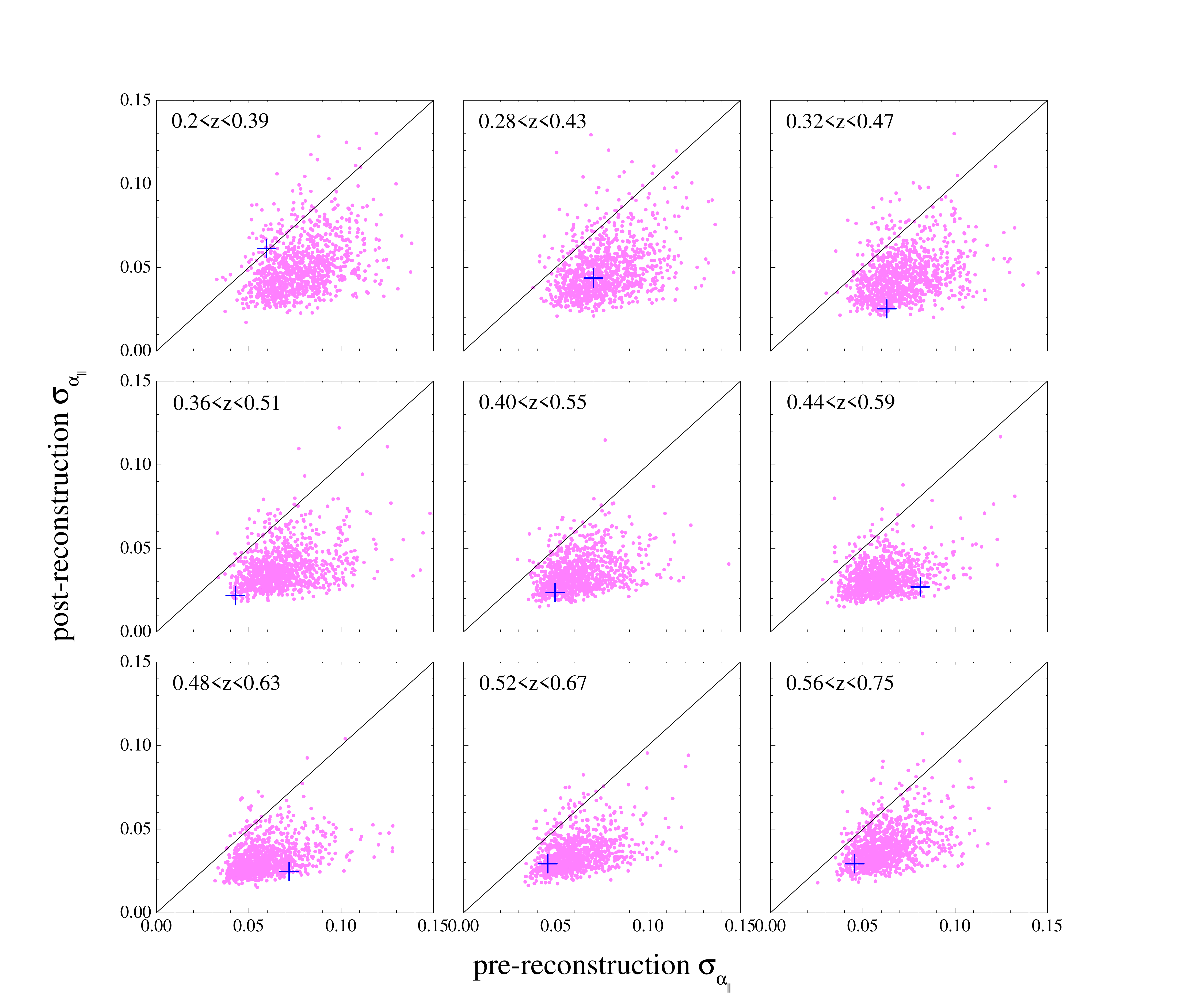}}
\caption{The scatter plots of errors of $\alpha_{\perp}$ (left panel) and $\alpha_{||}$ (right panel) using pre- and post reconstruction mock catalogue. Each magenta point denotes the $1\,\sigma$ error from each mock (totally 1000 mocks) and the cross (blue) is the error measured by data. }
\label{fig:aniso_mocktest}
\end{figure*}

Our recovered parameter values in nine redshift bins are well consistent with the input cosmology. For the isotropic results, we find that the greatest bias in $\alpha$ for the pre-reconstruction result is less than $0.4\%$, and is less than $0.2\%$ in the post-reconstruction case. The 1D distribution of the parameter $\alpha$ from mocks is shown in the histograms of Figure \ref{fig:iso_1D}, where the blue histograms are the pre-reconstruction results, and the red histograms are the post-reconstruction results. As expected, the BAO signals measured from the post-reconstructed mocks are more significant, as shown in the scatter plots of Figure \ref{fig:iso_mocktest}, where each point in the plot corresponds to the $1\,\sigma$ error value from each pre- and post-reconstructed mock. For the anisotropic results, on average the biases in the anisotropic parameters are less than $0.5\%$. We display the 1D distributions of the parameters, $\alpha_{\perp}$ and $\alpha_{\parallel}$ from mocks in the histograms of Figure \ref{fig:aniso_1D}. The scatter plots for the parameters, $\alpha_{\perp}$ and $\alpha_{\parallel}$ are shown in Figure \ref{fig:aniso_mocktest}.

\section{Results}
\label{sec:baoresult}

\subsection{Isotropic BAO measurements}
\label{sec:isobaoresult}

\begin{table}
\caption{The measurements on the isotropic BAO parameters and the reduced $\chi^2$ using the pre- and post-reconstruction catalogues, respectively.}
\centering
\begin{tabular}{cccc}
\hline\hline
 pre-reconstruction :                    &    &   &                                \\
$z_{\rm eff}$ &    $\alpha$ & $D_V/r_d$ & $\chi^2/\rm dof$  \\
0.31	&$	0.9916	\pm	0.0251	$&$	  8.31	\pm	0.21	$&	12.6/15	\\	
0.36	&$	0.9825	\pm	0.0320	$&$	  9.40	\pm	0.31	$&	13.2/15	\\	
0.40	&$	1.0000	\pm	0.0288	$&$	10.47	\pm	0.30	$&	20.9/15	\\	
0.44	&$	1.0155	\pm	0.0178	$&$	11.56	\pm	0.20	$&	16.5/15	\\	
0.48	&$	1.0234	\pm	0.0198	$&$	12.48	\pm	0.24	$&	22.3/15	\\	
0.52	&$	1.0074	\pm	0.0214	$&$	13.04	\pm	0.28	$&	21.9/15     \\	
0.56	&$	0.9924	\pm	0.0226	$&$	13.55	\pm	0.31	$&	22.1/15     \\	
0.59	&$	0.9906	\pm	0.0202	$&$	14.21	\pm	0.29	$&	21.0/15     \\	
0.64	&$	0.9770	\pm	0.0212	$&$	14.82	\pm	0.32	$&	15.1/15     \\
\hline  	
  post-reconstruction :               &     &     &                                \\
0.31	&$	0.9771	\pm	0.0172	$&$	  8.18	\pm	0.14	$&	16.8/15	\\	
0.36	&$	0.9925	\pm	0.0172	$&$	  9.50	\pm	0.16	$&	12.5/15	\\	
0.40	&$	1.0074	\pm	0.0149	$&$	10.54	\pm	0.16	$&	22.0/15	\\	
0.44	&$	1.0050	\pm	0.0116	$&$	11.44	\pm	0.13	$&	24.8/15	\\	
0.48	&$	1.0051	\pm	0.0109	$&$	12.26	\pm	0.13	$&	39.0/15	\\	
0.52	&$	0.9824	\pm	0.0108	$&$	12.72	\pm	0.14	$&	13.8/15	\\	
0.56	&$	0.9887	\pm	0.0112	$&$	13.50	\pm	0.15	$&	10.6/15	\\	
0.59	&$	0.9808	\pm	0.0141	$&$	14.07	\pm	0.20	$&	13.9/15	\\	
0.64	&$	0.9764	\pm	0.0159	$&$	14.81	\pm	0.24	$&	20.7/15	\\
\hline  \hline                       
\end{tabular}
\label{tab:iso_pre_post}
\end{table}

The correlation functions are measured with the bin width of 5 $\mpcoh$ bins, as shown in Figure \ref{fig:pre_NS_bins} and Figure \ref{fig:post_NS_bins}. We perform the fitting in the range $50 -150 \mpcoh$.

We present the constraints on the isotropic BAO scale in all redshift bins in Table \ref{tab:iso_pre_post}. Using the values of $D^{\rm fid}_V(z)/r^{\rm fid}_d$ for the fiducial cosmology, we derive the constraint on $D_V(z)/r_d$, as listed in the last two columns of Table \ref{tab:iso_pre_post}. The measurement precision on $D_V(z)/r_d$ from the pre-reconstruction catalogue can reach $1.8\%\sim3.3\%$. For the post-reconstruction, the precision is improved to be $1.1\%\sim1.8\%$. 

\begin{figure}
\centering
{\includegraphics[scale=0.32]{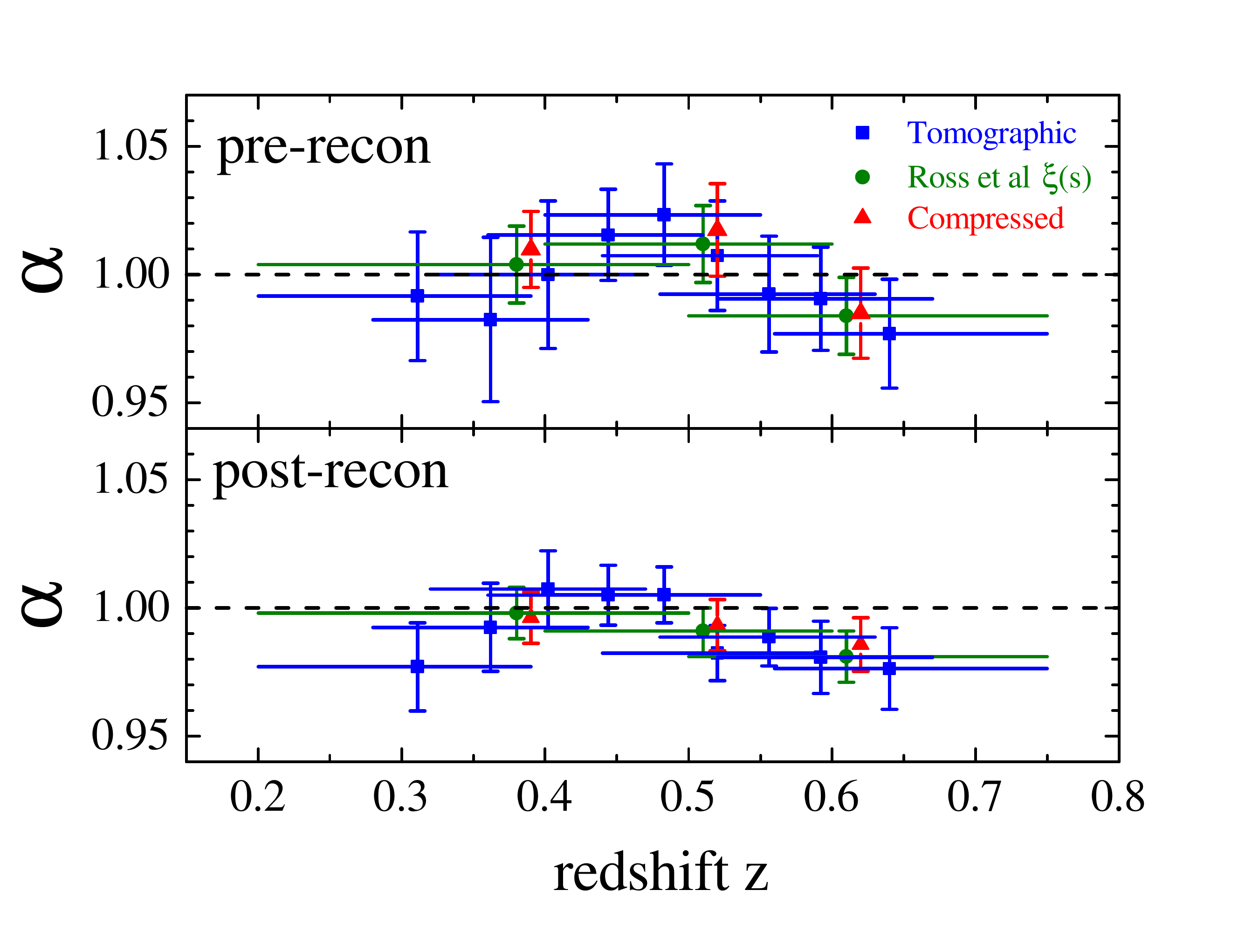}}
\caption{The fitting results on the isotropic BAO parameter, $\alpha$ using the pre- and post-reconstruction catalogues, respectively.}
\label{fig:NS_DV}
\end{figure}

The improvement on the measurement precision of $\alpha$ after reconstruction can be seen in Figure \ref{fig:NS_DV}, where we show our tomographic measurements in terms of the redshift in blue squares. The pre-reconstruction constraints are plotted in upper panel, and the lower panel shows the result after reconstruction. 

\begin{figure}
\centering
{\includegraphics[scale=0.3]{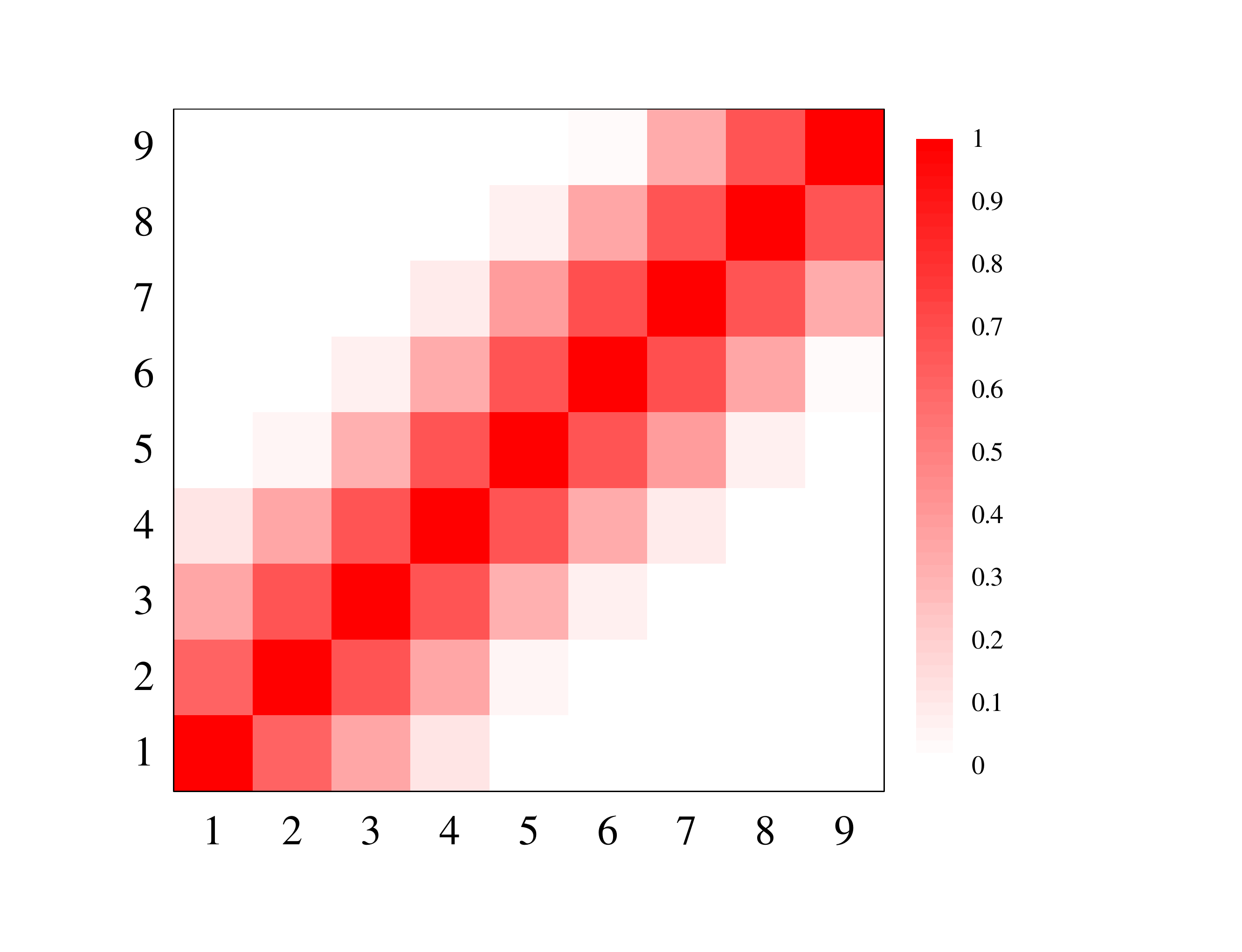}}
\caption{The normalized correlation of the parameters, $\alpha$, between different redshift slices.}
\label{fig:corr_iso_diffz_post}
\end{figure}

Since our redshift slices are highly correlated within the overlapping range, which is visualised in Figure \ref{fig:num_dis}, it is important to determine the correlations between redshift slices. We repeat the fitting on BAO parameter using each mock measurement, derive the covariance matrix between the $i$th $z$ bin and $j$th $z$ bin using $C_{ij}\equiv \langle \alpha_i \alpha_j \rangle- \langle \alpha_i \rangle \langle \alpha_j\rangle$, then calculate the correlation coefficient with $r_{ij}=C_{ij}/\sqrt{C_{ii}C_{jj}}$. The normalised correlations of $\alpha$ between redshift slices for the post-reconstruction are plotted in Figure \ref{fig:corr_iso_diffz_post}. It is seen that each bin is correlated to the 3 redshift bins next to it.

\subsection{Anisotropic BAO measurements}

\begin{table*}
\caption{The fitting results on the anisotropic BAO parameters, $\alpha_{\perp}$ and $\alpha_{\parallel}$, and their correlation coeffecient $r$ using the pre- and post-reconstruction catalogues, respectively.}
\begin{center} 
\begin{tabular}{ccccccccc}
\hline\hline 
                     &                              &  pre-reconstruction     &          &                     &                           & post-reconstruction        &       &             \\
$z_{\rm eff}$ &  $\alpha_{\perp}$ & $\alpha_{\parallel}$ &$r$&$\chi^2/\rm dof$ &$ \alpha_{\perp}$& $\alpha_{\parallel}$ &$r$& $\chi^2/\rm dof$ \\ \hline
0.31	&$	0.9596	\pm	0.0334	$&$	1.0378	\pm	0.0597	$&$	-0.40	$&$	26.0/30	$&$	0.9566	\pm	0.0212	$&$	1.0203	\pm	0.0614	$&$	-0.47	$&$	38.2/30	$\\
0.36	&$	0.9584	\pm	0.0334	$&$	1.0464	\pm	0.0704	$&$	-0.30	$&$	29.0/30	$&$	0.9762	\pm	0.0218	$&$	1.0275	\pm	0.0438	$&$	-0.36	$&$	35.3/30	$\\
0.40	&$	0.9706	\pm	0.0321	$&$	1.0414	\pm	0.0631	$&$	-0.27	$&$	41.7/30	$&$	0.9924	\pm	0.0200	$&$	1.0250	\pm	0.0253	$&$	-0.39	$&$	34.0/30	$\\
0.44	&$	0.9798	\pm	0.0228	$&$	1.0788	\pm	0.0426	$&$	-0.41	$&$	34.8/30	$&$	0.9971	\pm	0.0153	$&$	1.0168	\pm	0.0217	$&$	-0.36	$&$	27.5/30	$\\
0.48	&$	1.0104	\pm	0.0259	$&$	1.0341	\pm	0.0496	$&$	-0.42	$&$	38.0/30	$&$	1.0020	\pm	0.0130	$&$	1.0050	\pm	0.0235	$&$	-0.39	$&$	35.6/30	$\\
0.52	&$	1.0114	\pm	0.0272	$&$	0.9962	\pm	0.0810	$&$	-0.57	$&$	38.7/30	$&$	0.9935	\pm	0.0139	$&$	0.9560	\pm	0.0270	$&$	-0.49	$&$	12.1/30	$\\	
0.56	&$	1.0083	\pm	0.0276	$&$	0.9560	\pm	0.0718	$&$	-0.51	$&$	40.2/30	$&$	0.9878	\pm	0.0156	$&$	0.9877	\pm	0.0247	$&$	-0.43	$&$	16.1/30	$\\
0.59	&$	0.9926	\pm	0.0293	$&$	0.9982	\pm	0.0601	$&$	-0.53	$&$	37.3/30	$&$	0.9896	\pm	0.0180	$&$	0.9564	\pm	0.0307	$&$	-0.41	$&$	26.1/30	$\\	
0.64	&$	0.9656	\pm	0.0334	$&$	1.0014	\pm	0.0457	$&$	-0.43	$&$	25.7/30	$&$	0.9744	\pm	0.0219	$&$	0.9794	\pm	0.0294	$&$	-0.45	$&$	33.3/30	$\\	
\hline  \hline                       
\end{tabular}
\end{center}
\label{tab:aniso_pre_post_I}
\end{table*}

\begin{table}
\caption{The fitting results on the anisotropic BAO parameters, $D_A/r_d$  and $Hr_d$ using the pre- and post-reconstruction catalogues, respectively.}
\label{tab:aniso_pre_post_II}
\begin{center} 
\begin{tabular}{ccc}
\hline\hline 
pre-reconstruction: & & \\
$z_{\rm eff}$ &  $D_A/r_d$ & $Hr_d*10^3 [\rm km/s]$\\ 
0.31	&$	6.31	\pm	0.22	$&$	11.35	\pm	0.65	$ \\
0.36	&$	6.96	\pm	0.24	$&$	11.60	\pm	0.78	$ \\
0.40	&$	7.53	\pm	0.25	$&$	11.93	\pm	0.72	$ \\
0.44	&$	8.06	\pm	0.19	$&$	11.81	\pm	0.47	$ \\
0.48	&$	8.71	\pm	0.22	$&$	12.61	\pm	0.60	$ \\
0.52	&$	9.06	\pm	0.24	$&$	13.38	\pm	1.09	$ \\
0.56	&$	9.35	\pm	0.26	$&$	14.25	\pm	1.07	$ \\
0.59	&$	9.48	\pm	0.28	$&$	13.94	\pm	0.84	$ \\
0.64	&$	9.53	\pm	0.33	$&$	14.28	\pm	0.65	$ \\
 \hline
post-reconstruction: & & \\
$z_{\rm eff}$ &  $D_A/r_d$ &   $Hr_d*10^3 [\rm km/s]$  \\ 
0.31	&$	6.29	\pm	0.14	$&$	11.55	\pm	0.70	$ \\
0.36	&$	7.09	\pm	0.16	$&$	11.81	\pm	0.50	$ \\
0.40	&$	7.70	\pm	0.16	$&$	12.12	\pm	0.30	$ \\
0.44	&$	8.20	\pm	0.13	$&$	12.53	\pm	0.27	$ \\
0.48	&$	8.64	\pm	0.11	$&$	12.97	\pm	0.30	$ \\
0.52	&$	8.90	\pm	0.12	$&$	13.94	\pm	0.39	$ \\
0.56	&$	9.16	\pm	0.14	$&$	13.79	\pm	0.34	$ \\
0.59	&$	9.45	\pm	0.17	$&$	14.55	\pm	0.47	$ \\
0.64	&$	9.62	\pm	0.22	$&$	14.60	\pm	0.44	$ \\
\hline  \hline                       
\end{tabular}
\end{center}
\end{table}

We present the fitting result on the anisotropic BAO parameters in Table \ref{tab:aniso_pre_post_I} before and after reconstruction. Our measurements on $\alpha_{\perp}$ and $\alpha_{\parallel}$ are plotted in terms of redshift in blue squares of Figure \ref{fig:NS_DA} and \ref{fig:NS_H}, respectively. 

Based on the input fiducial values for $D^{\rm fid}_A/r^{\rm fid}_d$ and $H^{\rm fid}r^{\rm fid}_d$, we can obtain the constraints on the transverse and radial distance parameters, $D_A(z)/r_d$ and $H(z)r_d$, as listed in Table \ref{tab:aniso_pre_post_II}. The measurement precisions are within $2.3\%-3.5\%$ for $D_A(z)/r_d$ and $3.9\%-8.1\%$ for $H(z)r_d$ before the reconstruction. Using the reconstructed catalogues, the precisions are improved, which can reach $1.3\%-2.2\%$ for $D_A(z)/r_d$ and $2.1\%-6.0\%$ for $H(z)r_d$.

\begin{figure}
\centering
{\includegraphics[scale=0.3]{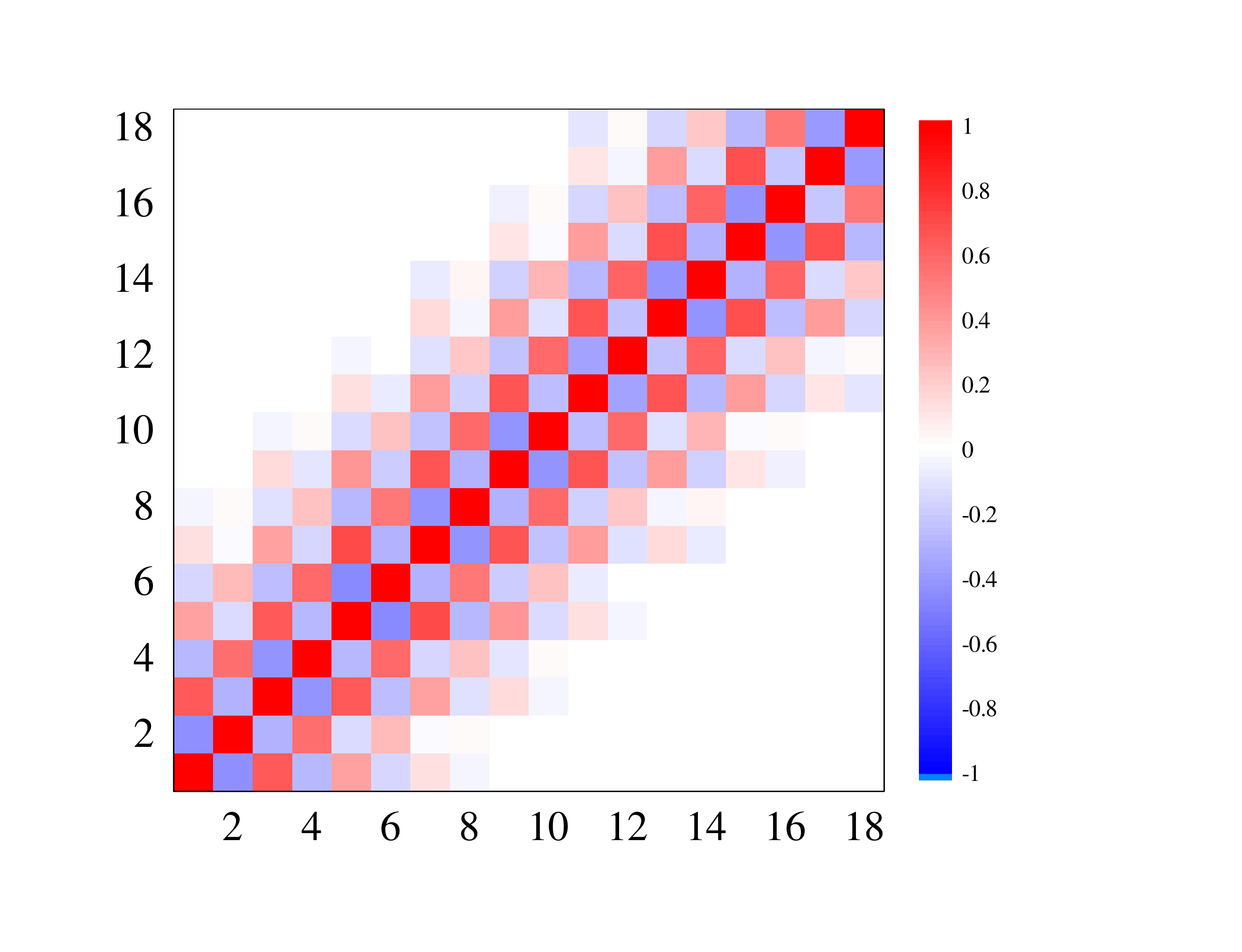}}
\caption{The normalised correlation of the parameters, $\alpha_{\perp}$  and $\alpha_{\parallel}$, between different redshift slices.}
\label{fig:corr_aniso_diffz_post}
\end{figure}

We determine the correlations between overlapping redshift slices using the measurements from mock catalogue. The calculation procedure has described in Section \ref{sec:isobaoresult}. The normalised correlated matrix of the parameters, $\alpha_{\perp}$  and $\alpha_{\parallel}$, between different redshift slices for the post-reconstruction are plotted in Figure \ref{fig:corr_aniso_diffz_post}.

\subsection{Result comparisons}

\begin{figure}
\centering
{\includegraphics[scale=0.28]{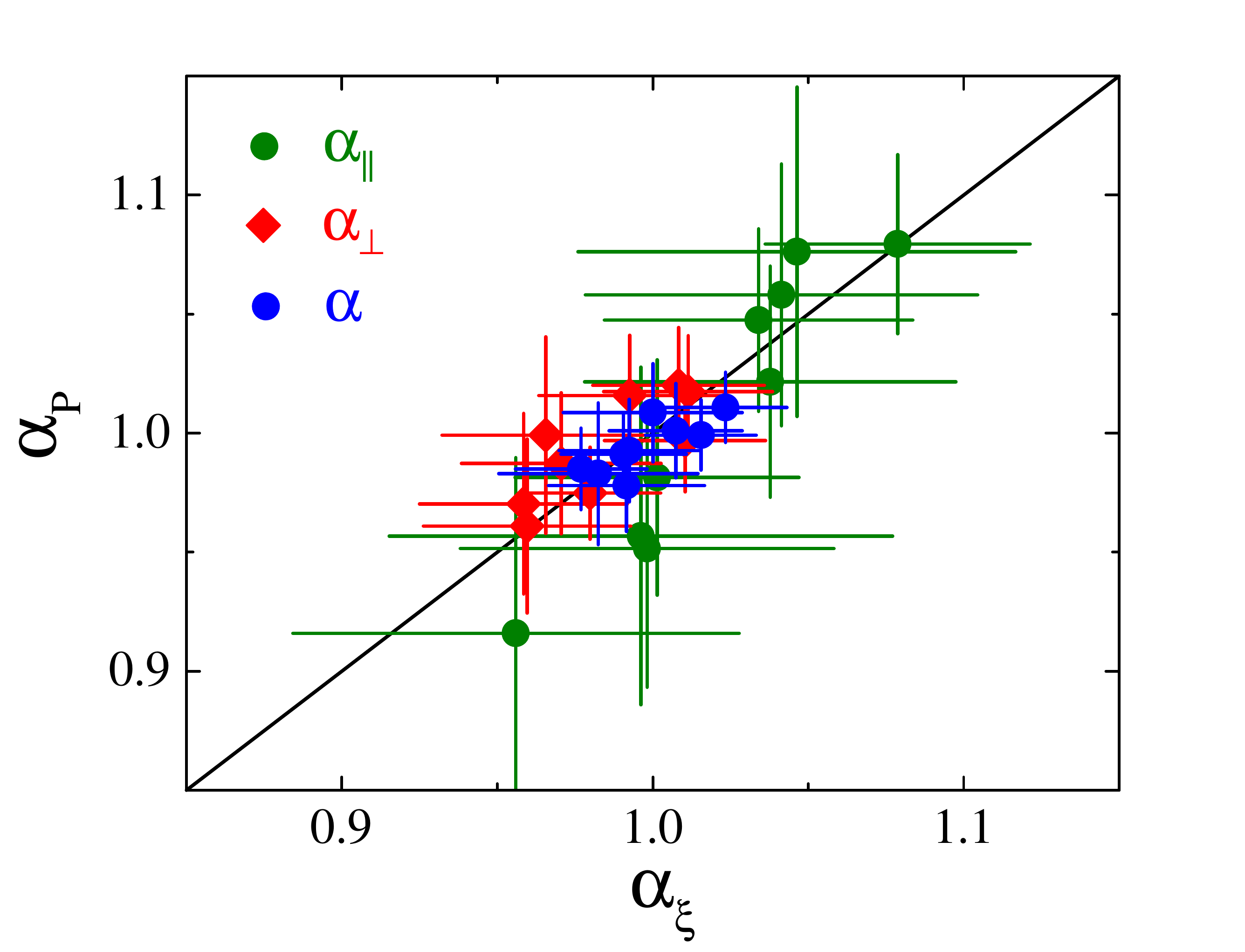}}
\caption{The comparison of our result on isotropic and anisotropic BAO parameters from the pre-reconstructed data with that in \citet{TomoPk}, measured in Fourier space.}
\label{fig:pk_vs_xi}
\end{figure}

We compare our pre-reconstructed results on the isotropic and anisotropic BAO parameters with the tomographic measurements using the power spectrum in Fourier space \citep{TomoPk}. The comparison is plotted in Figure \ref{fig:pk_vs_xi}. We can see that the isotropic results (blue points) agree well with each other. Because of the high correlations between anisotropic parameters, the comparison looks scattered, especially for the parameter $\alpha_{\|}$. Within the $1\,\sigma$ error, the results are consistent. The main difference is that \citet{TomoPk} use the monopole, quadrupole and hexadecapole in power spectrum, while we do not include the hexadecapole in our pre-reconstruction case. The role of the hexadecapole on anisotropic BAO constraints is discussed in detail \citep{TomoPk} . 

\begin{figure}
\centering
{\includegraphics[scale=0.3]{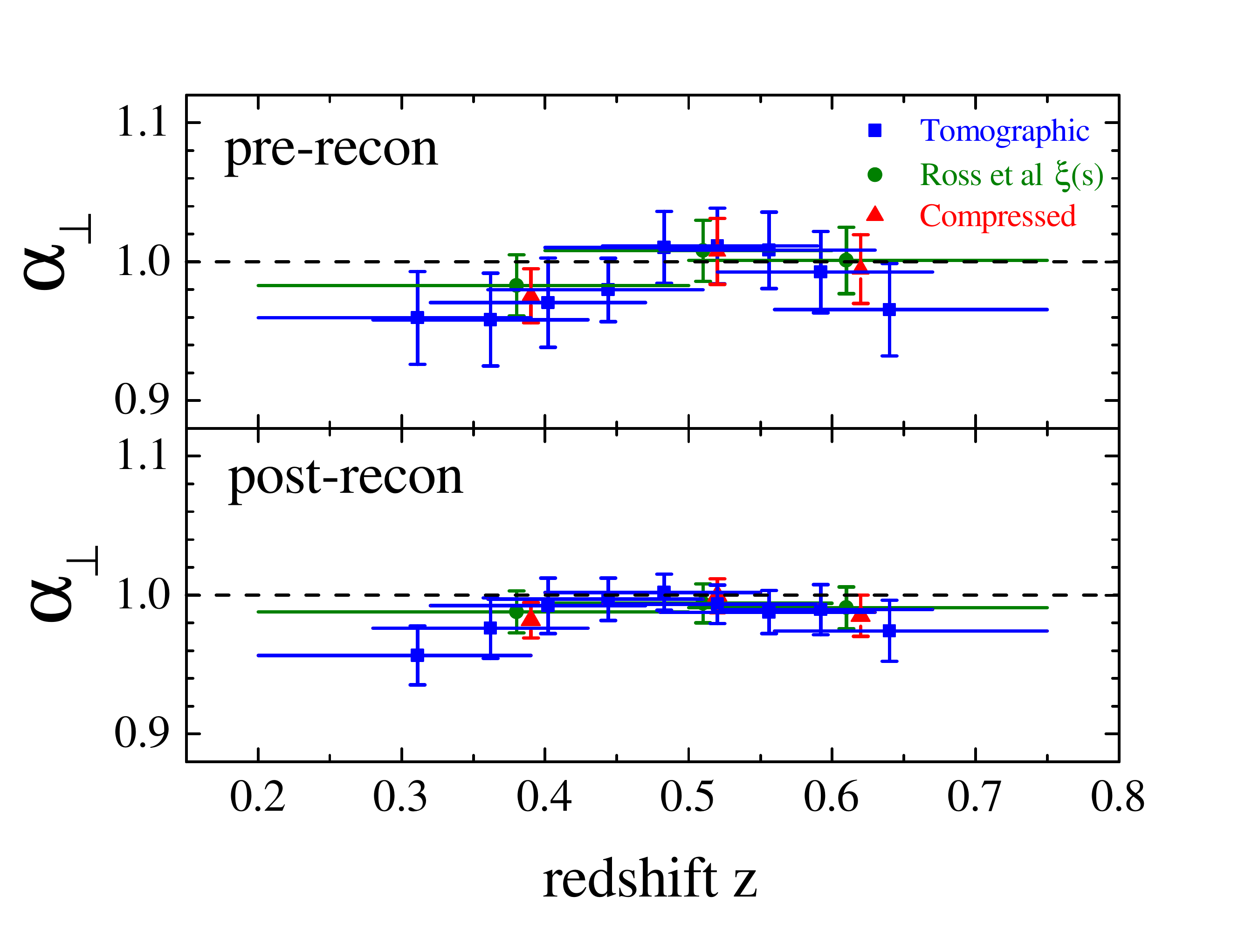}}
\caption{The fitting results on the anisotropic BAO parameter, $\alpha_{\perp}$ using the pre- and post-reconstruction catalogues, respectively.}
\label{fig:NS_DA}
\end{figure}

\begin{figure}
\centering
{\includegraphics[scale=0.3]{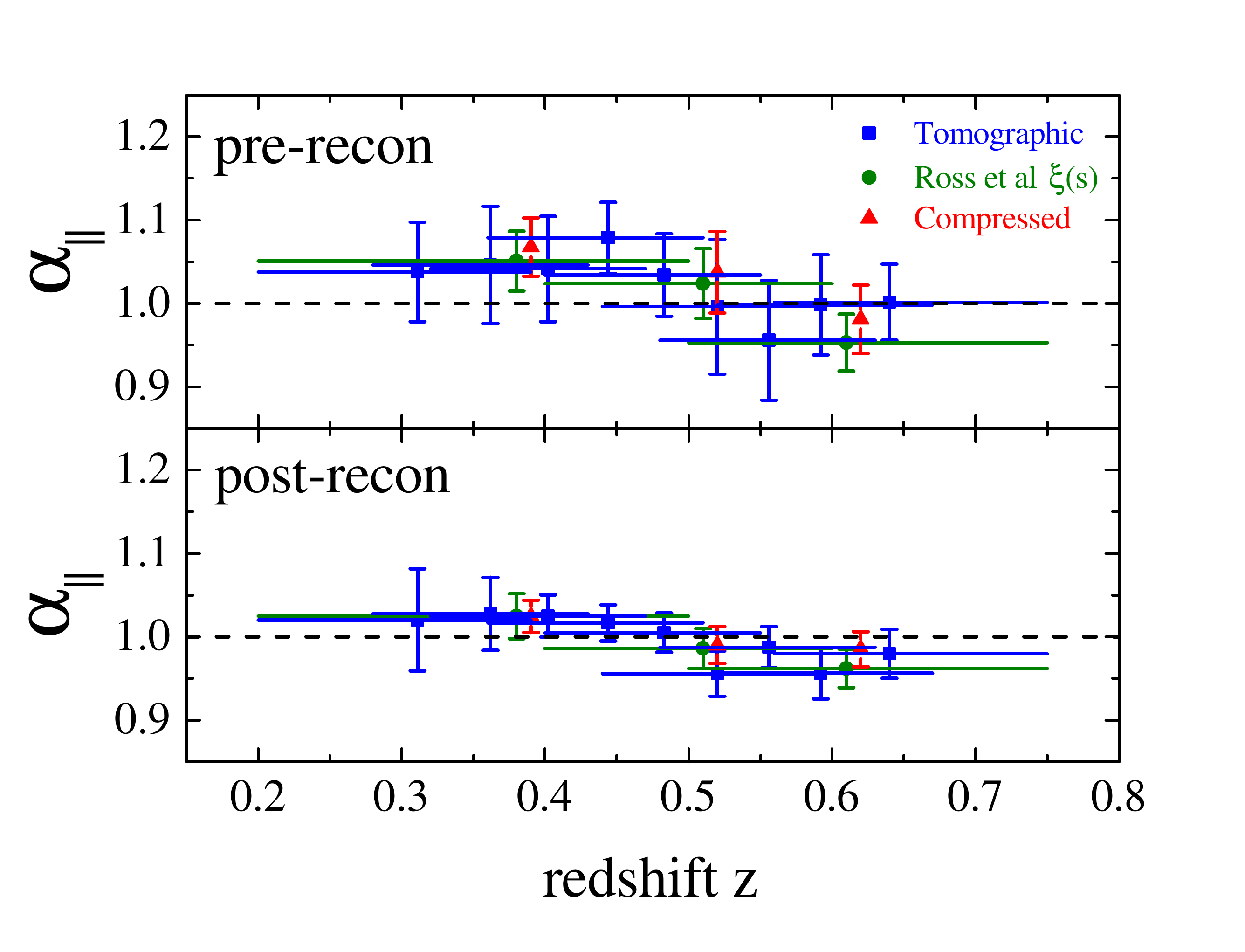}}
\caption{The fitting results on the anisotropic BAO parameter, $\alpha_{\|}$ using the pre- and post-reconstruction catalogues, respectively.}
\label{fig:NS_H}
\end{figure}

\begin{figure}
\centering
{\includegraphics[scale=0.28]{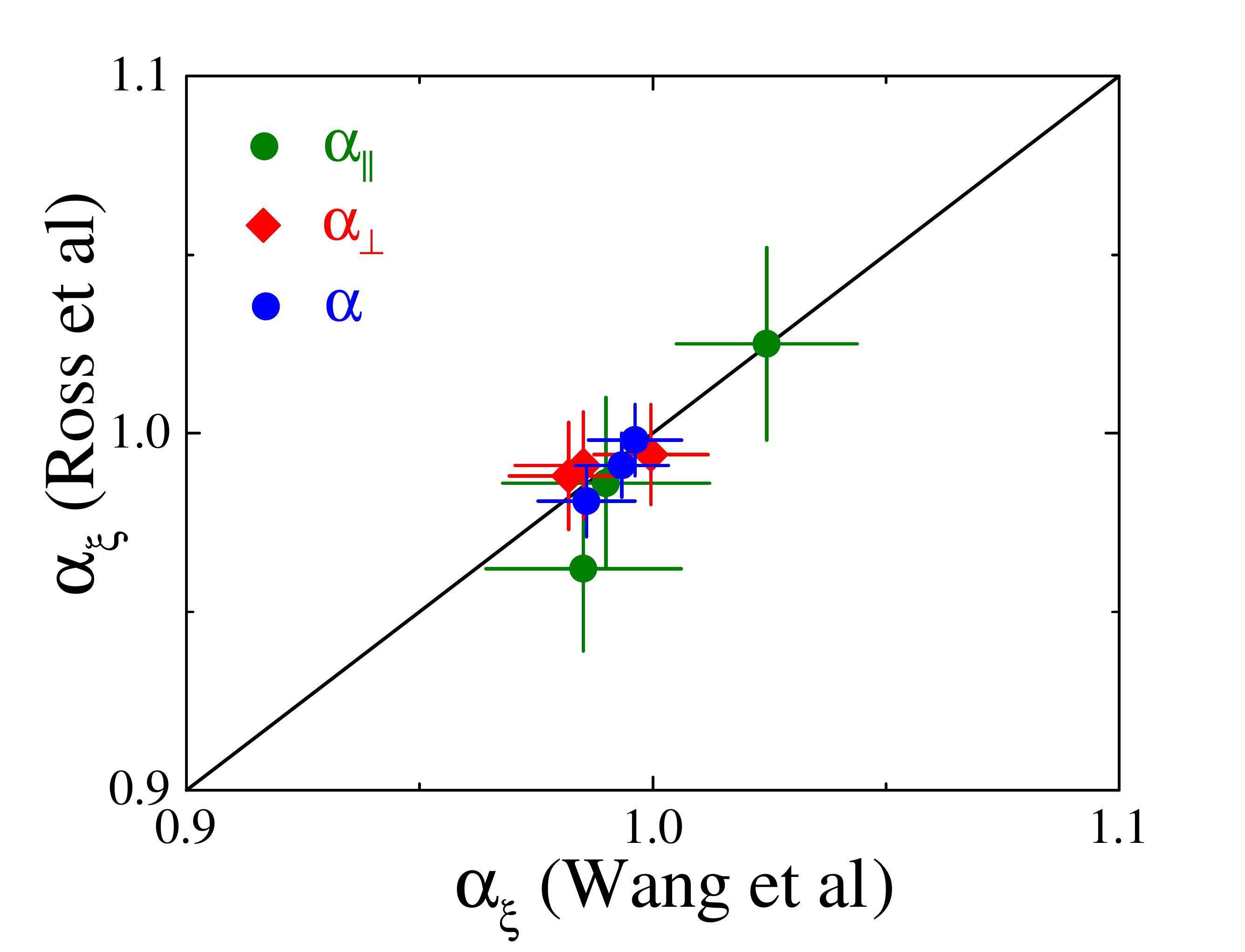}}
\caption{The comparison of our result on isotropic and anisotropic BAO parameters from the post-reconstructed data in the compressed 3 redshift bins with that in \citet{CF-sysweight}, also measured in configuration space.}
\label{fig:xi_vs_xi_3bin}
\end{figure}

In order to test the consistency between our measurements and the measurements in 3 redshift bins \citep{CF-sysweight}, we compressed our measurements into 3 redshift bins. Namely, we compressed the first 4 redshift bins, which covers the redshift range from 0.2 to 0.51, into one measurement. The compression is performed by introducing a parameter and fitting it to the measurements in these 4 redshift bins with their covariance matrix. The 5th and 6th bins ($0.4<z<0.59$) are compressed as the second measurement value. The last compressed measurement are from the remaining bins ($0.48<z<0.75$). The compression results are shown in red triangles of Figure \ref{fig:NS_DV}, \ref{fig:NS_DA} and \ref{fig:NS_H}. In these figures, the green points denote the results within 3 redshift bins from $\xi(s)$ measurements in \citet{CF-sysweight}, \ie \, two bins without overlapping between each other, [0.2, 0.5] and [0.5, 0.75], and an overlapping bin, [0.4, 0.6]. It is seen that with less redshift bins, more precise measurements and much tighter constraints can be obtained. In contrast, dividing more redshift bins in the tomographic case can capture the redshift information of galaxy clustering with more measurements at different effective redshifts. The comparison is plotted in Figure \ref{fig:xi_vs_xi_3bin}. We can see that our results are consistent with the measurements in \citet{CF-sysweight}. 

\begin{figure}
\centering
{\includegraphics[scale=0.28]{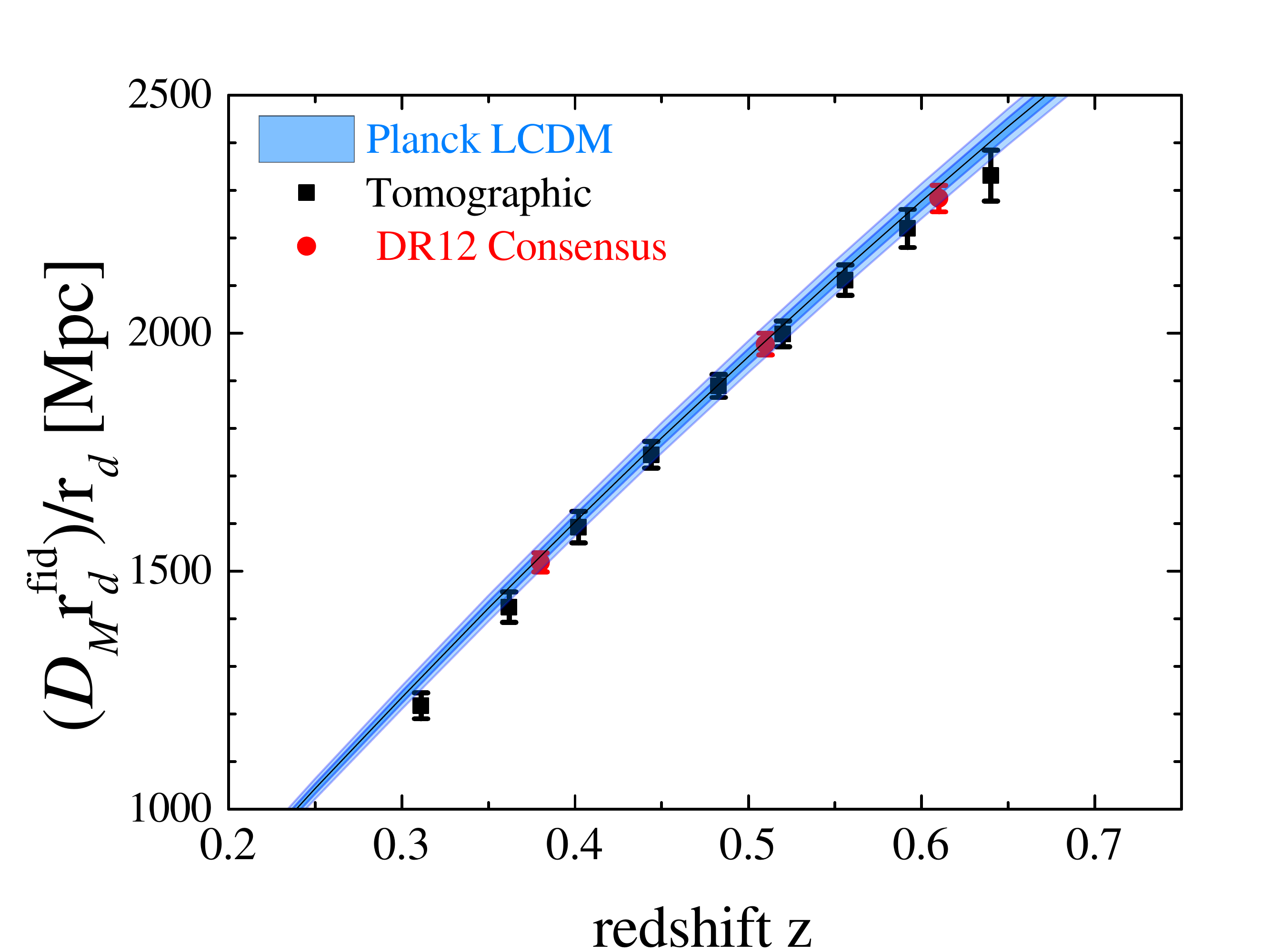}}
\caption{Our tomographic measurements on $D_Mr^{\rm fid}_d/r_d$ (black squares) in terms of redshift, compared with the consensus result (red points) in \citet{Alam2016} and the prediction from Planck assuming a $\Lambda$CDM model (blue bands). Here $D_M=(1+z)D_A$.}
\label{fig:PLC-DM_ov_rd}
\end{figure}

\begin{figure}
\centering
{\includegraphics[scale=0.28]{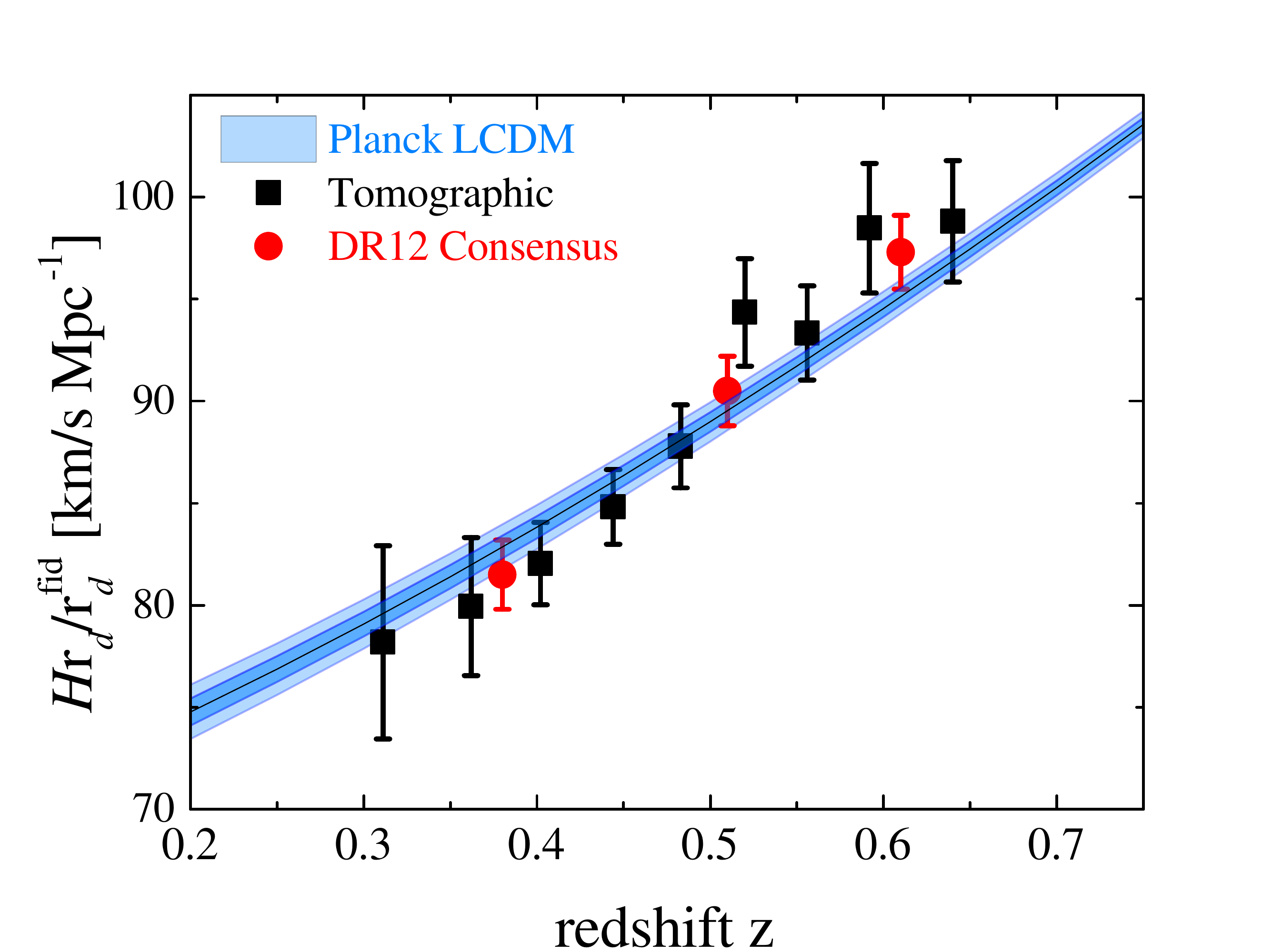}}
\caption{Our tomographic measurements on $H(z)r_d/r^{\rm fid}_d$ (black squares) in terms of redshift, compared with the consensus result (red points) in \citet{Alam2016} and the prediction from Planck assuming a $\Lambda$CDM model (blue bands).}
\label{fig:PLC-Hrd}
\end{figure}

The comparisons of our anisotropic BAO measurements with the three bins consensus measurements in \citet{Alam2016} are shown in Figure \ref{fig:PLC-DM_ov_rd} and \ref{fig:PLC-Hrd}, where the black squares are our measurements, and the red points are the consensus result, which are the combined constraints from the correlation function and power spectrum in \citep{Alam2016}. The blue bands correspond to the 68 and 95\% CL constraints in the $\Lambda$CDM using the Planck data assuming a $\Lambda$CDM model \citep{PLC2015likeli}. We can see these results are consistent.

\section{Constraints on cosmological models}
\label{sec:modelresult}

\begin{figure}
\centering
{\includegraphics[scale=0.28]{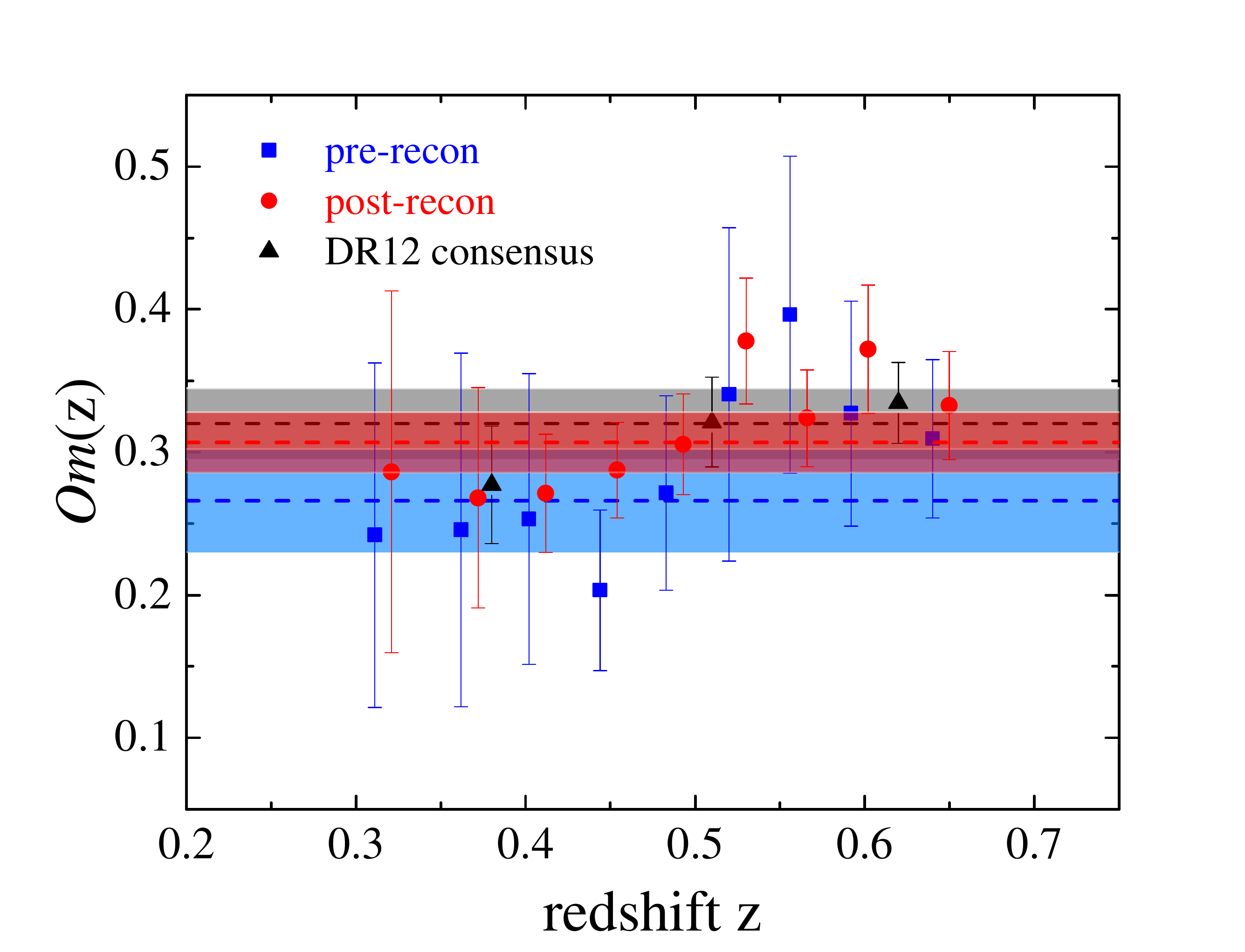}}
\caption{The $Om(z)$ values converted by our measurements on Hubble parameter in 9 redshift bins.}
\label{fig:om}
\end{figure}

Using our tomographic measurements on Hubble parameters, we do the $Om$ diagnostic, proposed by \citet{Sahni2008}. It is defined by the Hubble parameter
\ba
Om(z)\equiv\frac{[H(z)/H_0]^2-1}{(1+z)^3-1}.
\ea
In $\Lambda$CDM, $Om(z)=\Omega_m$. Using our measurements of $H(z)r_d$ and combining the fiducial values of $r_d=147.74 \,\rm Mpc$ and $H_0=67.8 \,\rm km/s/Mpc$, we convert our measurements to $Om(z)$, as shown in Figure \ref{fig:om}, where the blue squares are the pre-reconstruction tomographic measurements, the red points are the post-reconstruction tomographic measurements, and the black triangles are the consensus result in \citet{Alam2016}. 

To quantify the possible deviation from $\Lambda$CDM, we make a fit to the $Om(z)$ values with the covariance matrix between different redshift slices using a single parameter. As shown in Figure \ref{fig:om}, the black dashed line with the grey band are the best-fit value with $1\,\sigma$ error using the ``3 $z$bin" consensus result, the blue dashed line with the blue band are the ``9 $z$bin" pre-reconstruction tomographic result and the red dashed line with the red band are the ``9 $z$bin" post-reconstruction tomographic result. We obtain the fitting value, $\Omega_m=0.32\pm0.025$, with $\chi^2=1.73$ from the consensus result. Therefore, within $2\,\sigma$ regions there is no deviation from a constant $\Omega_m(z)$. Using our pre-reconstruction tomographic result, the fitting result is $\Omega_m=0.266\pm0.036$. The $Om(z)$ values in the pre-reconstruction case deviate from the fitting constant $\Omega_m(z)$ at about $2.01\,\sigma$ level. From our post-reconstruction result, the $Om(z)$ values deviate from the fitting result, $\Omega_m=0.307\pm0.021$, at about $2.78\,\sigma$.

We present the cosmological implications with our tomographic BAO measurements. We use the Cosmomc\footnote{\url{http://cosmologist.info/cosmomc/}}  \citep{cosmomc} code to perform the fittings on dark energy parameters in a time-varying dark energy with EoS, $w_{\rm DE}(a)=w_0+w_a(1-a)$ \citep{CPL2001, CPL2003}. 

We are using the combined data set, including the temperature and polarization power spectra from Planck 2015 data release \citep{PLC2015likeli}, the ``Joint Light-curve Aalysis" (JLA) sample of type Ia SNe \citep{SNeJLA}, the BOSS DR12 BAO distance measurements.  We compare the constraining power of different BAO measurements, \ie\, tomographic ``9 $z$bin" BAO measurements from the post-reconstructed catalogues, consensus ``3 $z$bin" measurements on BAO and RSD in \citet{Alam2016}, and the compressed ``1 $z$bin" BAO result from the post-reconstruction tomographic measurements.
 
The results of the parameters $w_0$ and $w_a$ are presented in Table \ref{tab:modelresult}. We can see the uncertainties of parameters are improved with the ``9 $z$bin" BAO measurements in our work.

\begin{figure}
\centering
{\includegraphics[scale=0.25]{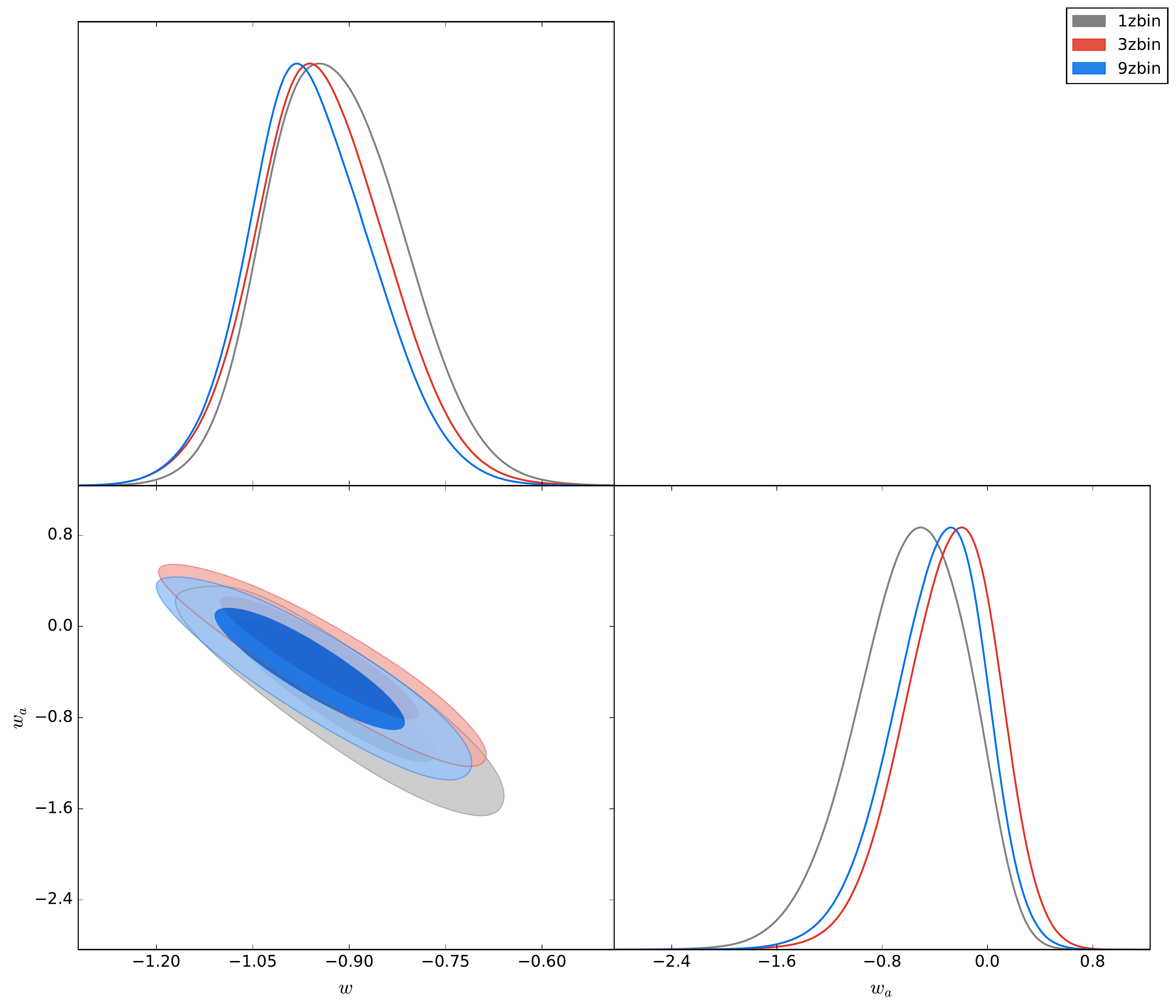}}
\caption{The 1D posterior distribution of $w$ and $w_a$ and their 2D contour plots in the CPL model from the compressed ``1 $z$bin" BAO (grey line and contour), consensus ``3 $z$bin" BAO and RSD (red line and contour), and tomographic ``9 $z$bin" BAO (blue line and contour).}
\label{fig:w0_wa_com}
\end{figure}

In $w_0w_a$CDM, comparing the tomographic ``9 $z$bin" with the non-tomographic ``1 $z$bin" results, the errors of $w_0$ and $w_a$ are improved by 6\% and 16\%, respectively. Using the Figure of Merit (FoM) \citep{FoM2009}, which is inversely proportional to the area of the contour as shown in Figure \ref{fig:w0_wa_com}, to quantify this improvement, the FoM is improved by a factor of 1.24 (FoM=49 for the grey contour from the ``1 $z$bin" result and FoM=61 for the blue contour from the ``9 $z$bin" result in Figure \ref{fig:w0_wa_com}). Comparing the ``9 $z$bin" with ``3 $z$bin" results, the ``9 $z$bin" BAO measurement give the slightly tighter constraints.

\begin{table}
\caption{Joint data constraints on dark energy EoS parameters $w_0$ and $w_a$ in the $w_0 w_a$CDM. Here we compare the constraining power of the BOSS DR12 BAO measurements, i.e. the tomographic ``9 $z$bin" measurements in this work, consensus ``3 $z$bin" measurements in \citet{Alam2016}, and the compressed ``1 $z$bin" result from tomographic measurements.}
\begin{center} 
\begin{tabular}{ccc}
\hline  \hline   
 Planck+JLA+BOSS  &   $w_0$ &$w_a$ \\  \hline
 Tomographic                 (9\,$z$bin) &  $-0.957\pm0.097$  &$-0.389\pm0.358$  \\ 
DR12 Consensus          (3\,$z$bin) &   $-0.942\pm0.101$  &$-0.288\pm0.359$  \\  
Compressed                 (1\,$z$bins) &   $-0.917\pm0.103$ &$-0.589\pm0.414$ \\ 
\hline  \hline                       
\end{tabular}
\end{center}
\label{tab:modelresult}
\end{table}

\section{Conclusion}
\label{sec:conclusion}

Measurements of the BAO distance scales have become a robust way to map the expansion history of the Universe. A precise BAO distance measurement at a single effective redshift can be achieved using the entire galaxies in the survey, covering a wide redshift range. However, the tomographic information is largely lost. To extract the redshift information from the samples, one possible way is to use overlapping redshift slices. 

Using the combined sample of BOSS DR12, we perform a tomographic baryon acoustic oscillations analysis using the two-point galaxy correlation function. We split the whole redshift range of sample, $0.2<z<0.75$, into multiple overlapping redshift slices, and measured correlation functions in all the bins. With the full covariance matrix calibrated using MultiDark-Patchy mock catalogues, we obtained the isotropic and anisotropic BAO measurements.

In the isotropic case, the measurement precision on $D_V(z)/r_d$ from the pre-reconstruction catalogue can reach $1.8\%\sim3.3\%$. For the post-reconstruction, the precision is improved, and becomes $1.1\%\sim1.8\%$. In the anisotropic case, the measurement precision is within $2.3\%-3.5\%$ for $D_A(z)/r_d$ and $3.9\%-8.1\%$ for $H(z)r_d$ before the reconstruction. Using the reconstructed catalogues, the precision is improved, which can reach $1.3\%-2.2\%$ for $D_A(z)/r_d$ and $2.1\%-6.0\%$ for $H(z)r_d$.

We present the comparison of our measurements with that in a companion paper \citep{TomoPk}, where the tomographic BAO is measured using multipole power spectrum in Fourier space. We find an agreement within the $1\,\sigma$ confidence level. The derived 3-bin results from our tomographic measurements are also compared to the 3-bin measurements in \citet{CF-sysweight}, and a consistency is found. 

We perform cosmological constraints using the tomographic 9-bin BAO measurements, the consensus 3-bin BAO and RSD measurements, and the compressed 1-bin BAO measurement. Comparing the constraints on $w_0w_a$CDM from 9-bin and 1-bin BAO distance measurements, the uncertainties of the parameters, $w_0$ and $w_a$ are improved by 6\% and 16\%, respectively. The dark energy FoM is improved by a factor of 1.24. Comparing the ``9 $z$bin" with ``3 $z$bin" results, the ``9 $z$bin" BAO measurement give the slightly tighter constraints.

The future galaxy surveys will cover a larger and larger cosmic volume, and there is rich tomographic information in redshifts to be extracted. The method developed in this work can be easily applied to the upcoming galaxy surveys and the gain in the temporal information is expected to be more significant. 
       
\section*{Acknowledgements}

YW is supported by the NSFC grant No. 11403034. GBZ and YW are supported by National Astronomical Observatories, Chinese Academy of Sciences, and by University of Portsmouth. 

Funding for SDSS-III has been provided by the Alfred P. Sloan Foundation, the Participating Institutions, the National Science Foundation, and the US Department of Energy Office of Science. The SDSS-III web site is \url{http://www.sdss3.org/}. SDSS-III is managed by the Astrophysical Research Consortium for the Participating Institutions of the SDSS-III Collaboration including the University of Arizona, the Brazilian Participation Group, Brookhaven National Laboratory, Carnegie Mellon University, University of Florida, the French Participation Group, the German Participation Group, Harvard University, the Instituto de Astrofisica de Canarias, the Michigan State/Notre Dame/JINA Participation Group, Johns Hopkins University, Lawrence Berkeley National Laboratory, Max Planck Institute for Astrophysics, Max Planck Institute for Extraterrestrial Physics, New Mexico State University, New York University, Ohio State University, Pennsylvania State University, University of Portsmouth, Princeton University, the Spanish Participation Group, University of Tokyo, University of Utah, Vanderbilt University, University of Virginia, University of Washington, and Yale University.

This research used resources of the National Energy Research Scientific Computing Center, which is supported by the Office of Science of the U.S. Department of Energy under Contract No. DE-AC02-05CH11231, the SCIAMA cluster supported by University of Portsmouth, and the ZEN cluster supported by NAOC.

\bibliographystyle{mn2e}
\bibliography{tomoBAOpk}

\label{lastpage}

\end{document}